\documentclass[amsmath,amssymb,nofootinbib]{revtex4-1}
\usepackage{ae}
\usepackage[T1]{fontenc}
\usepackage[ansinew]{inputenc}
\usepackage{amsmath}
\usepackage{amssymb}
\usepackage{color}
\usepackage{graphicx}
\usepackage{subfigure}
\usepackage{longtable}
 \usepackage[usenames,dvipsnames]{pstricks}
\usepackage{epsfig}
\usepackage{pst-grad}
\usepackage{pst-plot}

\newcommand{\va}{\scriptscriptstyle}

\newcommand{\CS}{C_{\va \Sigma}}

\newcommand{\R}{\mathbb{R}}

\DeclareFontFamily{U}{rsfs}{}                                                                                          %         Formal Script            %
\DeclareFontShape{U}{rsfs}{m}{n}{<5> rsfs5 <6><7> rsfs7          %
  <8><9><10><10.95><12><14.4><17.28><20.74><24.88> rsfs10}{}     %
\DeclareMathAlphabet{\mathfs}{U}{rsfs}{m}{n}                     %
\newcommand{\mfs}[1]{\mathfs {#1}}                               %
\newcommand{\sH}{{\mfs H}}
\newcommand{\sL}{{\mfs L}}

\newcommand{\sM}{{\mfs M}}

\newcommand{\sO}{{\mfs O}}

\newcommand{\Hp}{{\sH}_{phys}}

\newcommand{\Hk}{{\sH}_{kin}}

\newcommand{\Ha}{{\sH}_{aux}}

\newcommand{\inter}{{\lrcorner}}

\newcommand{\su}{\mathfrak{su}}

\newcommand{\Tr}{\mathrm{Tr}}
\newcommand{\Irrep}{\mathrm{Irrep}}

\newcommand{\be}{\nopagebreak[3]\begin{equation}}
\newcommand{\ee}{\end{equation}}
\newcommand{\bee}{\nopagebreak[3]\begin{equation*}}
\newcommand{\eee}{\end{equation*}}
\newcommand{\ba}{\nopagebreak[3]\begin{eqnarray}}
\newcommand{\ea}{\end{eqnarray}}
\newcommand{\baa}{\nopagebreak[3]\begin{eqnarray*}}
\newcommand{\eaa}{\end{eqnarray*}}
\newcommand{\la}{\label}
\newcommand{\n}{\nonumber}

\begin{document}
\scalebox{1}

\title{$2+1$ gravity with positive cosmological constant in LQG:\\
a proposal for the physical state}

\date{\today}

\author{Daniele Pranzetti}
\affiliation{Centre de Physique Th\'eorique,
Universit\`e de Provence, Aix-Marseille I,\\
 Campus de Luminy, 13288
Marseille, France.}

\begin{abstract}
In this paper, I investigate the possible 
quantization, in the context of LQG, of three dimensional gravity in the case of positive cosmological constant $\Lambda$
and try to make contact with alternative quantization approaches already existing in the literature. Due to the appearance of an anomaly in the constraints algebra, previously studied as a first step of the analysis, alternative techniques developed for the quantization of systems with constraints algebras not associated with a structure Lie group need to be adopted. Therefore, I introduce an ansatz for a physical state which gives some transition amplitudes in agreement with what one would expect from the Turaev-Viro model. Moreover, in order to check that this state implements the right dynamicss, I show that it annihilates the master constraint for the theory up to the first order in $\Lambda$.
\end{abstract}
\maketitle

\section{introduction}\label{Intro}
\pagestyle{plain}

Three dimensional quantum gravity represents an interesting example of completely integrable system and can be defined in several different ways (see \cite{Carlip} for a comprehensive review). The first model to appear in the literature in the late `60$s$ was the Ponzano-Regge state sum model \cite{Ponzano-Regge}  for 3-dimensional Euclidean quantum gravity without cosmological constant using the Lie group $SU(2)$. The Ponzano-Regge model defines a partition function for a triangulated compact 3-manifold by assigning an irreducible representation of $SU(2)$ to each edge of the triangulation, and a certain weight to each interior edge, triangle, tetrahedron. By means of $SU(2)$ recoupling theory, after summing over all possible values of the spin on every edge in the interior of the manifold, one doesn't always get a finite result due to the infinite dimension of the set of irreducible $SU(2)$ representations\footnote{One can show that this divergences are due to infinite contributions of pure gauge modes and eliminate them by appropriate gauge fixing. This procedure boils down to working with those triangulations where infinite sums are not present \cite{freidel-louapre}.}.

Several years later, a regularized version of the Ponzano-Regge model was introduced by Turaev and Viro \cite{Turaev-Viro}. In defining the new state sum, the two authors replaced the Lie group $SU(2)$ with its quantum deformation $U_q SL(2)$ and when the deformation parameter $q$ is a $r$-th root of unity, then there are only a finite number of irreducible representations, thus always providing a finite answer. 

Shortly after, in his two seminal works \cite{Witten}, Witten showed how Chern-Simons theory was closely related to three dimensional gravity, by providing a quantization of the latter through the definition of a Chern-Simons path integral. Witten also proved that the Turaev-Viro state sum was equivalent to a Feynman path integral with the Chern-Simons action for the group product $SU(2)_k \otimes SU(2)_{-k}$, where $k$ is the level of the Cehern-Simons theory which is related to the level $r$ of the Turaev-Viro model by $k=r-2$, showing in this way the connection between the Turaev-Viro model and three dimensional quantum gravity with cosmological constant $\Lambda$, once the relation $k^2 = 4\pi^2/\Lambda$ holds \cite{Ooguri-Sasakura} (see also \cite{Freidel} for the connection between Turaev-Viro model and gravity). Quantum groups also enter the quantization of Chern-Simons theory in the so-called combinatorial quantization approach (\cite{Combinatorial Quantization}), where a quantum deformation of the structure group is introduced as an intermediate regularization.

In the context of Loop Quantum Gravity (LQG) \cite{LQG}, only the case of vanishing cosmological constant is clearly understood. The quantization is in this case a direct implementation of Dirac's quantization program for gauge systems. The basic unconstrained phase space variables are represented as operators in an auxiliary Hilbert space (or kinematical Hilbert space $\Hk$ spanned by spin network states) where the constraints are represented by {\em regularized} quantum operators.  A nice feature of the regularization (which is both natural but also unavoidable in the context of LQG) is that it leads to regulated quantum constraints satisfying the appropriate quantum constraints algebra. There is no anomaly, i.e. the constraints close a Lie algebra. As shown in \cite{Noui Perez}, this feature together with the background independent nature of the whole treatment allows to define a regularization of the formal expression for the generalized projection operator into the kernel of curvature constraint introduced in \cite{Reisenberger, Rovelli}. One can, therefore, build the physical Hilbert space of the theory  through this precise definition of the physical inner product which can be represented as a sum over spin foams (\cite{Foam}) whose amplitudes coincide with those of the Ponzano-Regge model. For the connection between the LQG program and the combinatorial quantization
formalism approach to the quantization of three dimensional gravity in the case of vanishing cosmological constant see \cite{Meusburger:2008bs}.

In the case of non-vanishing cosmological constant, one would then expect to be able to understand the Turaev-Viro amplitudes as the physical transition amplitudes or physical inner product between kinematical spin network states. In this paper we move the first steps into this direction. The LQG treatment of the non-vanishing cosmological constant case was started in \cite{Anomaly}, where the authors concentrated on the very basic starting point of the Dirac program: the study of the quantization of the constraints and their associated constraints algebra. The results of \cite{Anomaly} show the appearance, proper to the nature of the kind of regularizations admitted by the LQG mathematical framework (\cite{Perez}), of a quantization anomaly in the constraints algebra, namely the Lie algebra structure is broken. 

In order to make contact with the Turaev-Viro model, we expect that the implementation of the dynamics will enable us to ``construct'' the quantum group structure starting from a kinematical Hilbert space where no deformation of the classical gauge Lie group is introduced at all (see \cite{Smolin} for an alternative approach in which a quantum deformation is introduced by hand at the kinematical level). In this direction, in this paper I attempt to contour the obstacle found in the $\Lambda\neq0$ case, namely the breaking of the Lie algebra structure of the theory constraints, by means of some alternative formulations developed for the quantization of systems with constraints algebras which are not associated with a structure Lie group. More specifically, I introduce an ansatz for a physical state which gives some transition amplitudes in agreement with what one would expect from the Turaev-Viro model and, in order to check that this state implements the right dynamics, I apply the master constraint program \cite{Thiemann:2003zv}. The results of this paper represent a first step towards the speculated, but still unproven, match, as perfectly realized in the $\Lambda=0$ case, between the covariant and canonical approaches to the problem of $2+1$ quantum gravity in the presence of a non-vanishinng cosmological constant.

The paper is organized as follows.
In Section \ref{Constraints Algebra} I recall the classical constraints algebra  and the analysis of its quantization by briefly recalling the regularization prescription introduced in \cite{Anomaly}.
In section \ref{Quantum Dim} I introduce an ansatz for a physical state which gives some transition amplitudes in agreement with what one would expect from the Turaev-Viro model and in section \ref{Master Constraint} I apply the master constraint technique to prove that the ansatz previously introduced implements, up to the first order in $\Lambda$, the right dynamics. In section \ref{Conclusions} conclusions are presented.

\section{Phase space, gauge symmetries and constraints algebra}\label{Constraints Algebra}

\subsection{Classical Analysis}\label{Classical Analysis}

We are interested in (Euclidean) three dimensional gravity with positive cosmological
constant in the first order formalism. The space-time $\sM$ is
a three dimensional oriented smooth manifold and the action is
given by
\begin{equation}\label{eq:action}
S[e,\omega]=\int_{\sM}\Tr[e\wedge
F(\omega)+\frac{\Lambda}{3} e \wedge e\wedge e],
\end{equation}
\noindent where $e$ is a $\su(2)$ Lie algebra valued
$1$-form, $F(\omega)$ is the curvature of the three dimensional
connection $\omega$ and $\Tr$ denotes a Killing form on
$\su(2)$. We assume the space time topology to be $\sM= \Sigma × \R$ where $\Sigma $ is a Riemann surface of arbitrary genus.

Upon the standard 2+1 decomposition, the phase space is parametrized by the
pull back to $\Sigma$ of $\omega$ and $e$. In local coordinates we
can express them in terms of the $2$-dimensional connection
$A^i_a$ and the triad field $E^b_j = \epsilon^{bc} e^k_c
\eta_{jk}$ where $a = 1, 2$ are space coordinate indices and $i, j
= 1, 2, 3$ are $\su(2)$ indices and $\epsilon^{ab}=-\epsilon^{ba}$ with $\epsilon^{12}=1$ (similarly $\epsilon_{ab}=-\epsilon_{ba}$ with $\epsilon_{12}=1$). The Poisson bracket among these variables is given  by
\begin{equation}\label{eq:symplectic}
\{A^i_a(x), E^b_j(y)\}=\delta^b_a \delta^i_j \delta^{(2)}(x,y).
\end{equation}
Due to the underlying $SU(2)$ and diffeomorphisms gauge invariance
the phase space variables are not independent and satisfy the following set of first class constraints. The
first one is the analog of the familiar Gauss law of Yang-Mills theory, namely
\begin{equation}\label{eq:kinematic}
G_i\equiv D_a E^a_i=0,
\end{equation}
\noindent where $D_a$ is the covariant derivative with
respect to the connection $A$. The constraint
(\ref{eq:kinematic}) is called the Gauss constraint. It encodes the condition that the connection be
torsion-less and it generates infinitesimal $SU(2)$ gauge
transformation. The second constraint reads
\begin{eqnarray}\label{eq:dynamics}
\nonumber C^i &=&\epsilon^{ab} (F_{ab}^i(A)+\Lambda \epsilon^{i}_{\ jk}e^j_a e^k_b)=0\\ &=& \epsilon^{ab} F_{ab}^i(A)+\Lambda \epsilon_{cd}\epsilon^{ijk}E^c_j E^d_k=0,
\end{eqnarray} where in the first line we have written the constraint in terms of the triad field, while in the second line we have used the electric field.
This second set of first class constraints generate local `translations'. As shown below, 
diffeomorphisms invariance of three dimensional gravity is associated to these two previous sets of constraints, i.e. diffeomorphisms can be written as linear combinations of the transformations generated by (\ref{eq:kinematic}) and (\ref{eq:dynamics}).
In order to exhibit the underlying (infinite dimensional) gauge symmetry Lie algebra it is convenient to smear  the constraints (\ref{eq:dynamics}) and (\ref{eq:kinematic}) with arbitrary test fields $\alpha$ and $N$, which we assume not depending on the phase space variables, they read:
\begin{equation}\label{eq:kinematic_smeared}
G(\alpha)=\int_\Sigma \alpha^i G_i=\int_\Sigma \alpha^i D_a E^a_i=0
\end{equation}
and
\be\label{eq:dynamics_smeared} C(N)=\int_\Sigma N_i C^i=\int_\Sigma N_i(F^i(A)+\Lambda \epsilon^{ijk}E_j E_k)=0.
\ee
The constraints algebra is then
\begin{eqnarray}\label{eq:constraints_algebra}
\nonumber \{C(N),C(M)\}&=& \Lambda \ G([N,M])\\ 
\nonumber \{G(\alpha),G(\beta)\}&=&G([\alpha,\beta])\\ 
\{C(N),G(\alpha)\}&=&C([N,\alpha]),
\end{eqnarray}
where $[a,b]^i=\epsilon^{i}_{\ jk} a^jb^k$ is the commutator of $\su(2)$. 

%For future use it will be convenient to split the constraint $C(N)$ as
%\be
%C(N)=F(N)+E( \Lambda N) \label{splitting1},
%\ee
%where
%\be\label{splitting2}
%F(N)=\int_\Sigma N_i(F^i(A)), \ \ \ \ \ \ \ \ \ \ \ \ \ E(\Lambda N)=\int_\Sigma \Lambda \epsilon^{ijk} N_i E_j E_k.
%\ee
The transformations generated by the Gauss constraint $G(\alpha)$ and the curvature constraint $C(N)$ read
\ba\label{eq:variations}
\delta_\alpha A^i&=&\{A^i,G(\alpha)\}=(d_A\alpha)^i~~~~~~~~~~~~~~~~\delta_\alpha E_i=\{E_i,G(\alpha)\}=-\epsilon_{ijk}\alpha^jE^k\n\\
\delta_N A^i&=&\{A^i,C(N)\}=-2\Lambda \epsilon^{ijk}N_j E_k~~~~~~\delta_N E_i=\{E_i,C(N)\}=(d_A N)_i.
\ea
Provided that the $E$-field be non-degenerate, on shell, diffeomorphisms generated by a vector field $v$ can be written as linear combinations of the previous transformations with parameters $\alpha^i(v)=v \inter A^i=v^aA^i_a$ and $N_i(v)=v \inter E_i=\epsilon_{ab}v^aE_i^b$, namely
\be
\sL_v A^i=\delta_{\alpha(v)}A^i+\delta_{N(v)}A^i~~~~~~~~
\sL_v E_i=\delta_{\alpha(v)}E_i+\delta_{N(v)}E_i,
\ee
where $\sL_v$ is the Lie derivative operator along the vector field $v$.

\subsection{Quantum Analysis}\label{Quantum Analysis}

\begin{figure} \centerline{\hspace{0.5cm}\(
\begin{array}{ccc}
\includegraphics[height=4cm]{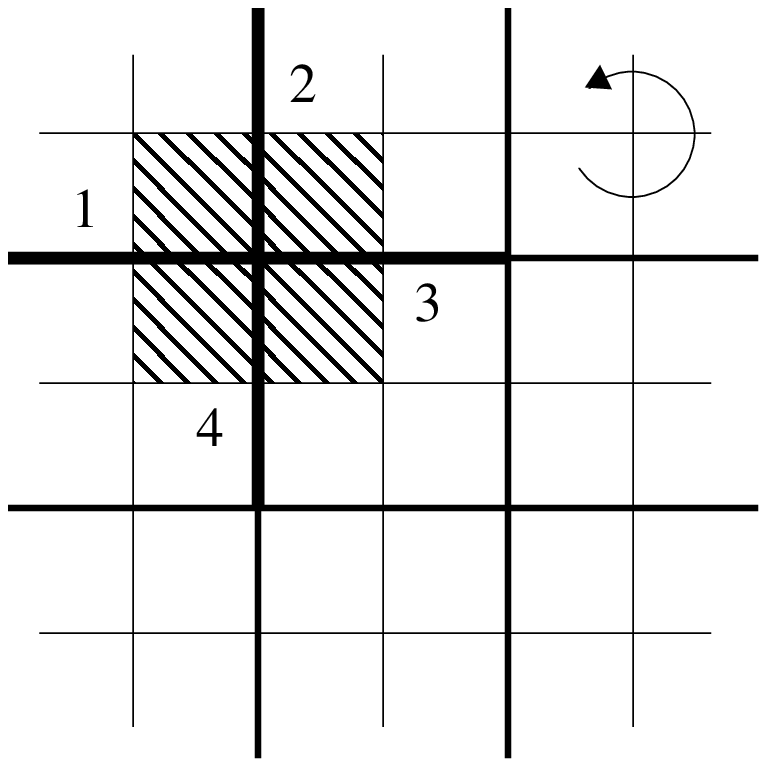}
\end{array}\ \ \ \ \ \ \ \ \
\begin{array}{ccc}
\includegraphics[height=1cm]{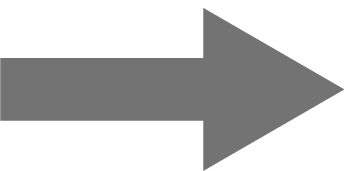}
\end{array} \ \ \ \ \ \ \ \ \
\begin{array}{ccc}
\includegraphics[height=4cm]{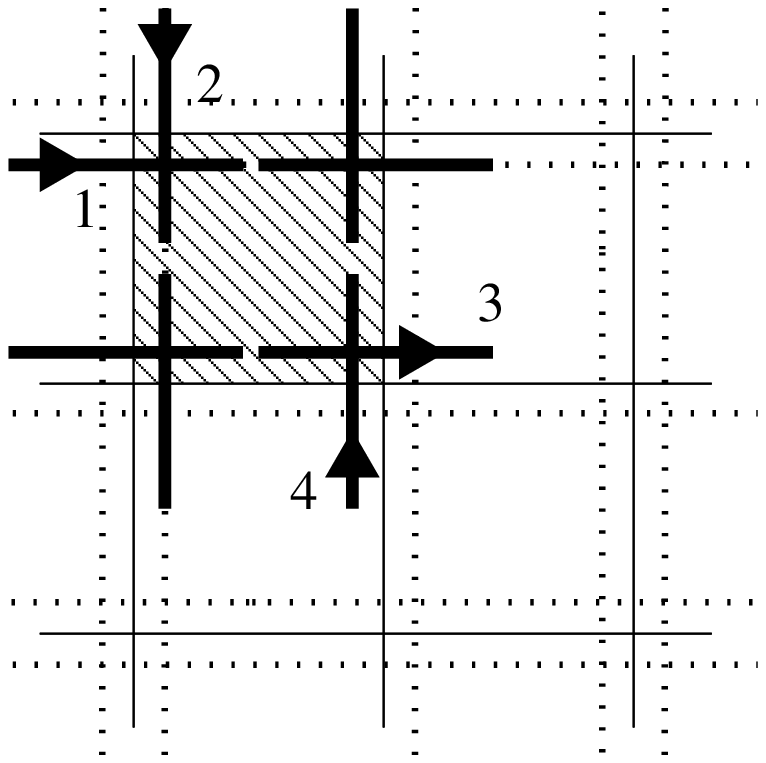}
\end{array}
\)} \caption{On the left: portion of the cellular decomposition $\CS$ (thin lines) and its dual $\CS^{\star}$ (thick lines). On the right: the edges of $\CS^{\star}$ are shifted toward the corresponding nodes. The flux operators necessary for the definition of the regularization of $E[\Lambda N]$ are defined in terms of the latter shifted dual edges---in this way, their action is consistent with that of the Gauss constraint operator as shown in \cite{Anomaly}.}
\label{fig:Cellular_decomposition}
\end{figure}

In order to provide a quantization of the constraints (\ref{eq:kinematic_smeared})-(\ref{eq:dynamics_smeared}), we first have to translate the
classical variables entering their definition, namely the connection $A$ and the electric field $E$, in terms of holonomies of the connection and fluxes of the electric field.
In order to do that we need to define a discrete structure on top of which we can construct these extended variables. We do so by introducing  an
arbitrary finite cellular decomposition $\CS$ of $\Sigma$.  We denote $n$  the number of plaquettes
(2-cells) which from now on will be denoted by the index $p\in \CS$. We assume the plaquettes to be squares  with edges  (1-cells denoted $e\in \CS$) of  length $\varepsilon$ in a local coordinate system. It will also be necessary to use the dual complex $C_{\Sigma^{\va *}}$ with its dual plaquettes $p^{\va *}\in C_{\Sigma^{\va *}}$ and edges $e^{\va *}\in C_{\Sigma^{\va *}}$ (see FIG. \ref{fig:Cellular_decomposition}). Both cellular decompositions inherit the orientation from the orientation of $\Sigma$.
The cellular decomposition defines the regulating structure. We now need to write the classical constraints (\ref{eq:kinematic_smeared})-(\ref{eq:dynamics_smeared})  in terms of extended variables in such a way that the naive continuum limit is satisfied.

The phase space variables $E^a_i$ and $A_a^i$  are discretized as follows:
the local connection $A_a^i$ field is now replaced by the assignment of group elements $h(e)=P \exp(-\int_e A)\in SU(2)$ to the set of edges $e\in \CS$.
We discretize the triad field $E^a_i$ by assigning to each dual
1-cell $e^{\va *}$ the $\emph{su(2)}$ element $E_i({e^{\va *}})\equiv \int_{e^{\va *}} \epsilon_{ab} E_i^b(x) dx^a$, i.e. the flux of electric field across the dual edge $e^{\va *}$.

After having discretized our phase space variables we can now quantize them.
The (generalized) connection is quantized  by promoting the holonomy to an operator acting by multiplication on a state $\Psi[A]\in Cyl$, where $Cyl$ is the space of cylindrical functionals (\cite{ash3}), of the auxiliary Hilbert space $\Ha$ of the theory spanned by spin network states as follows:
\be
\widehat{h_\gamma[A]} \triangleright \Psi[A] \; = \; h_\gamma[A] \Psi[A].
\label{ggcc}
\ee
The triad is associated with operators in $\Ha$ defining the flux of electric field across one dimensional lines which can be defined from its action on holonomies:
\ba
\widehat{E_i({e^{\va *}})}\triangleright h_{\gamma}[A]=\frac{i}{2}s_p \left\{\begin{array}{ccc} o_{e^{\va *}\gamma} \ \tau_i h_{\gamma}[A]\ \ \ \ \mbox{(for $e^{\va *}$ target of $\gamma$)}\\ o_{e^{\va *}\gamma}\ h_{\gamma}[A]\tau_i \ \ \ \ \mbox{(for $e^{\va *}$ source of $\gamma$)} \end{array}\right., \label{fluxx}
\ea
where the curve $\gamma$ is assumed to have one of its endpoints at $e^{\va *}$, $o_{e^{\va *}\gamma}=\pm1$ is the sign of the orientation of the pair of oriented curves in the order $(e^{\va *},\gamma)$, and
where $s_p=\hbar G$ is the Planck length in three dimensions (the action vanishes if the curves are tangential to each other).

With the decomposition of $\Sigma$ introduced above, we can now write the regularized  versions of the constraints (\ref{eq:kinematic_smeared}) and (\ref{eq:dynamics_smeared}). Following the prescription introduced in \cite{Anomaly}, we have:
\begin{equation}\label{eq:kinematic_discretized}
G^{\va R}(\alpha)=\sum_{p^{\va *}\in C_{\Sigma^{\va *}}} {\rm tr}[\alpha^{p^{\va *}} G^{p^{\va *}}]=0
\end{equation}
and
\begin{equation}\label{eq:dynamics_discretized}
C^{\va R}(N)=\sum_{p\in\CS} {\rm tr}[N^p C^p]=0,
\end{equation}
where $G^{p^{\va *}}$ and $C^p$ are explicitly defined in \cite{Anomaly}.

Finally, the allowed states will be a subset $Cyl (\CS)\subset Cyl$
consisting of all cylindrical functions whose underlying graph is contained in the one-skeleton
of $\CS$. In other words, the allowed graphs must consist of collections of $1$-cells  $e\in\CS$.

With this prescription, the quantum version of the constraints algebra (\ref{eq:constraints_algebra}) of gravity in $2+1$ dimensions with non-vanishing cosmological constant reads \cite{Anomaly}:
\begin{eqnarray}\label{eq:quantum constraints_algebra}
&&[C^{\va R}(N),C^{\va R}(M)]= {\Lambda} \,G^{\va R}(\frac{{\rm tr}[W]}{2}[N,M]) \nonumber\\
&&[G^{\va R}(N),G^{\va R}(M)]= G^{\va R}([N,M])\n\\ 
&&[C^{\va R}(N),G^{\va R}(M)]=  C^{\va R}([N,M]).
\end{eqnarray}
Relations (\ref{eq:quantum constraints_algebra}) show that just the commutators among the scalar constraints present an anomaly due to the presence of the factor $\frac{{\rm tr}[W]}{2}$ in the smearing of the
Gauss law on the r.h.s. of the first equation. We see that the regularization does not break the internal gauge group $SU(2)$; however, it does break the part of the gauge symmetry group related to spacetime
 diffeomorphisms.  Notice also that the anomaly is a genuine quantum effect. If we had computed the Poisson algebra of regularized constraints instead we would have found basically the same result (where commutators are replaced by Poisson brackets). However, in that case the problematic factor disappears in the continuum limit as $\frac{{\rm tr}[W]}{2}=1+\sO(\epsilon^4)$.

This anomaly found in \cite{Anomaly} represents an unexpected difficulty for the implementation of the standard Dirac quantization in the LQG representation. In particular, the group averaging techniques used in the context of $2+1$ gravity without cosmological constant to solve the first class constraints at the quantum level are not viable. Nevertheless, from the point of view of $3+1$ gravity, the anomaly appearing in the constraints algebra is a mild one. In fact, in the four dimensional case one has to deal with structure functions in the constraints algebra already at the classical level. From this perspective the problem is well known and studied. Therefore, in order to contour this obstacle, one can try to follow some alternative formulations developed for the quantization of systems with constraints algebras which are not associated with a structure Lie group such as the master constraint \cite{Thiemann:2003zv}. That's what I am going to do in the next sections, where I first introduce an ansatz for a physical state which gives some transition amplitudes in agreement with the recoupling theory of the quantum group $U_q SL(2)$ and then I use the master constraint technique to show that this state solves the curvature constraint in the case $\Lambda \neq 0$, up to the first order in $\Lambda$. In this way I'll try to provide evidence to the conjecture that the anomaly found in \cite{Anomaly} may be at the end related with the deformation of the classical symmetry of gravity leading to the quantum group structure underlying the quantization of 2+1 gravity with non-vanishing cosmological constant found by other methods. 

\section{Quantum dimension}\label{Quantum Dim}

In order to deal with the difficulties related to the appearance of an anomaly in the constraint algebra, as illustrated in the previous section, we are now going to introduce an ansatz for the physical state of $2+1$ gravity with positive cosmological constant and then use it to compute the physical transition amplitude between the vacuum and a Wilson loop state. Since, as stressed in the previous section, each state in the theory has to be built on top of the regulating structure introduced to discretize the fundamental variables of the theory---in order to be able to use the LQG framework for quantization---, our ansatz has to be defined over the cellular decomposition $\CS$ of $\Sigma$. The removal of the regulator consists of taking the limit $n\rightarrow \infty$, which correspond to refining more and more the cellular decomposition\footnote{However, in the removal of the regulator, the smooth configuration where each plaquette shrinks to a single point in space can never be reached in the framework of LQG since the fundamental variables of the theory to be quantized are of extended nature (holonomies along  edges and flux operators across dual edges):  in LQG it doesn't exists such a thing as point-like operators. Moreover, this is the origin of the anomaly found in the quantum constraints algebra (\ref{eq:quantum constraints_algebra}): the fact that loop operators never shrink to a single point in LQG prevents from getting rid of the term ${\rm tr}[W]/2$ in the first commutator and, therefore, the algebra to close.}. More precisely, the ansatz for the physical state we want to study in the rest of the paper reads: 
\begin{eqnarray}\la{eq:Ansatz}
\Psi =\frac{1}{2}(\Psi_+ + \Psi_-)&=&\lim_{n\rightarrow \infty}\frac{1}{2}\Bigg(\prod_p \sum_{j_p}(2j_p+1)\left(1+i\frac{\sqrt{\Lambda}}{n}(2j_p+1)\right)\chi_{j_p}(W_p)\nonumber\\
&+&\prod_p \sum_{j_p}(2j_p+1)\left(1-i\frac{\sqrt{\Lambda}}{n}(2j_p+1)\right)\chi_{j_p}(W_p)\Bigg),
\end{eqnarray}
where the discrete structure used to define the state $\Psi$ is of the same nature of the one introduced above to regularize the constraints. In the previous expression, $\chi_j$ denotes the trace of the $j$-representation matrix of $g\in SU(2)$ and $W_p$ is the holonomy around the plaquette $p$. 
The state (\ref{eq:Ansatz}) is non-normalizable. This might represent a problem at first, however, as it is well known in LQG, physical states lie outside $\Ha$. The physical Hilbert space $\Hp$ is not a subspace of $\Ha$ and physical states correspond to `distributional states' in $Cyl^*$, the dual of the dense set $Cyl\subset \Ha$. Through the inner product we can identify $|\phi>$ of a spin network basis in $Cyl$ to a dual spin network $<\phi|\in Cyl^*$. Non-normalizability, for instance, is also a feature of the projection operator into the kernel of the curvature constraint defined in \cite{Noui Perez} in the case of $\Lambda=0$ in order to implement the dynamics of gravity---notice that the state (\ref{eq:Ansatz}) exactly reduces to the this projection operator in the limit $\Lambda\rightarrow 0$. Even if this operator is not well defined in $\Ha$, it provides a well defined expression for the physical inner product which allows to recover the Ponzano-Regge model.

\begin{figure}[h!]
\centering
  \includegraphics[width=5.7cm,angle=360]{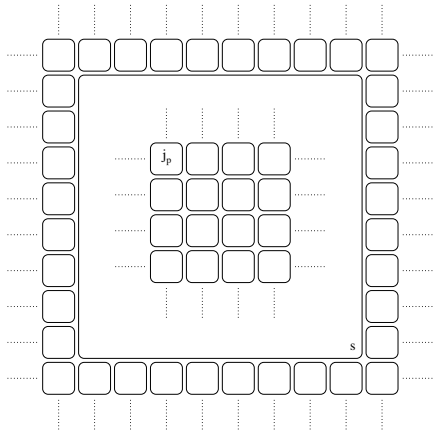}
   \caption{For a given n there are finitely many plaquettes  inside and outside the loop $\alpha$.}
   \label{Sovra}
\end{figure}

Let us now compute the scalar product defined by the Ashtekar-Lewandowski measure (\cite{ash3}) between the state introduced above and a generic loop state in the $\Irrep$ $s$. In the following we will assume $\Lambda>0$ and $\Sigma$ to have the topology of a sphere, i.e. Euler characteristic $\chi=2$.
We'll do it first for the first part of the state $\Psi$ and then in an analogous way for the second one, namely
\begin{equation}
<s|\Psi_+>=\!\int\!\! \left(\!\prod_h dg_h\!\right) \chi_s(g_\alpha)\!\lim_{n\rightarrow \infty}\prod_p \sum_{j_p}(2j_p+1)\!\left(\!1+i\frac{\sqrt{\Lambda}}{n}(2j_p+1)\!\right)\!\chi_{j_p}(W_p),
\end{equation}
where $g_\alpha$ is the holonomy around the loop $|s>$.
The previous scalar product is graphically represented if FIG. \ref{Sovra},
where each plaquette has a given representation $j_p$ assigned. If we denote by $n_I$ and $n_O$ the number of plaquettes inside and outside the loop $|s>$ respectively ($n=n_I +n_O$), one can easily perform the integration over the group elements associated to the internal and external (to the loop $|s>$) edges which are not along the border of $|s>$ by use of the relation
\be\label{eq:integration}
\begin{array}{c}  \includegraphics[width=1cm,angle=360]{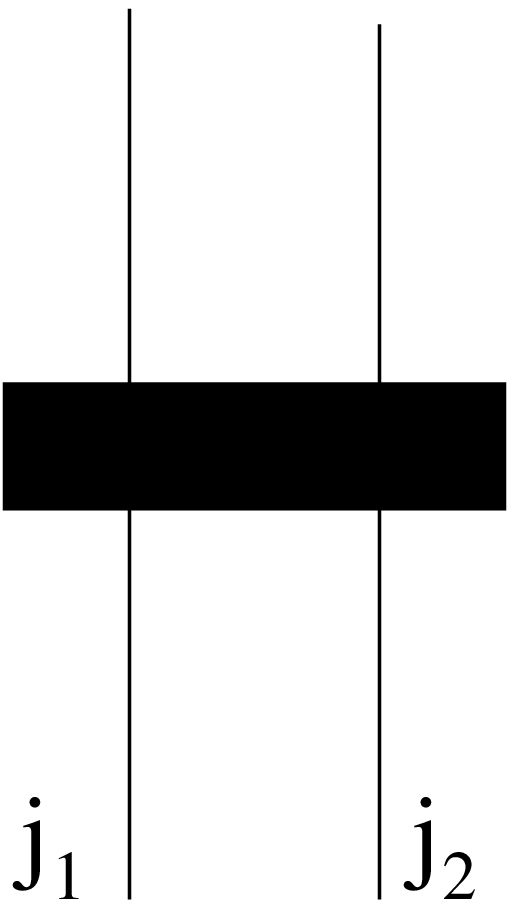}\end{array}=\frac{1}{(2j_1+1)}\delta_{j_1 j_2}\begin{array}{c}  \includegraphics[width=1cm,angle=360]{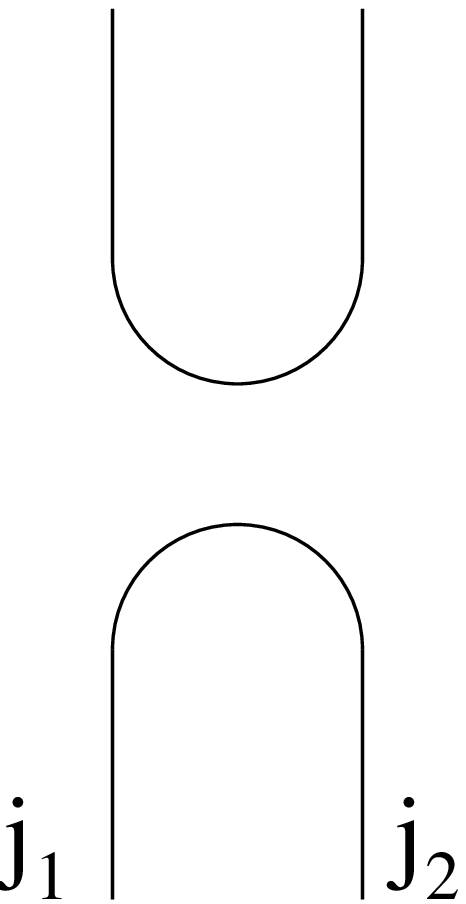},\end{array}
\ee
where the black box symbolize integration over the group element in common between the two edges. If we do so, we see that all the internal plaquettes are forced to be in the same spin representation, which we call $j$, and so do all the external ones, which we call $k$. Therefore after integration on these group elements one is left with the following expression for the scalar product:
\baa
<\!s|\Psi_+\!>&=&\!\int\!\! \left(\prod_{\tilde{h}} dg_{\tilde{h}}\right)\!\!\lim_{n\rightarrow \infty}\Big[\sum_k (2k+1)\!\left(1+i\frac{\sqrt{\Lambda}}{n}(2k+1)\right)^{n_O}\sum_j(2j+1)\!\left(1+i\frac{\sqrt{\Lambda}}{n}(2j+1)\right)^{n_I}\Big]
\begin{array}{c}  \includegraphics[width=2cm,angle=360]{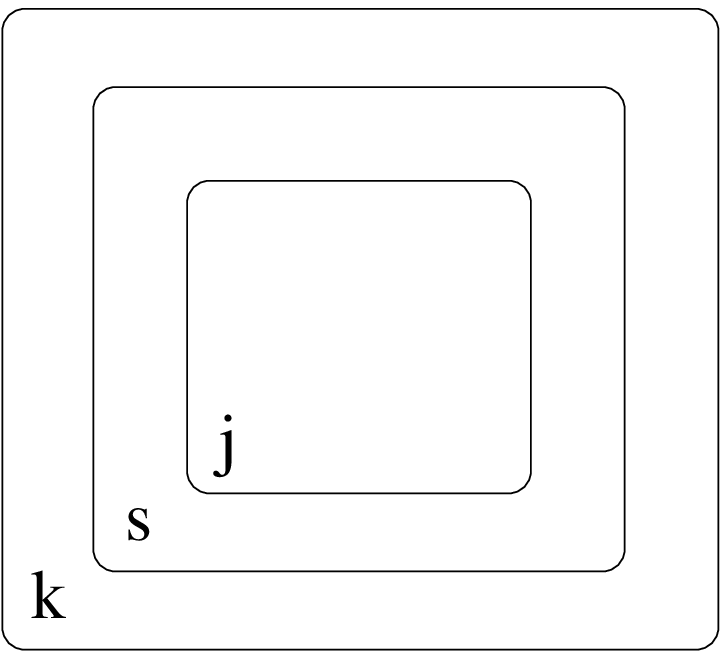}\end{array},
\eaa
%\begin{figure}
 % \includegraphics[width=3cm,angle=360]{Squares}
%\end{figure}
where the integrals left in the previous expression are only on the group elements $g_{\tilde{h}}$ associated to the edges along the border of $|s>$. It is now easy to see that, performing these last integrals, one obtains a theta net which evaluates to one but imposes some admissibility condition between the spins $j$, $k$ and $s$. Thereby, performing all the integrals and imposing the limit $n\rightarrow \infty$ by keeping $n_O$ fixed and sending $n_I$ to infinity, one is left with
\ba
<s|\Psi_+>&=&\sum_{k=0}^{\infty} (2k+1)e^{i\frac{n_O}{n}\sqrt{\Lambda}(2k+1)}
\sum_{j=|k-s|}^{k+s}(2j+1)e^{i\sqrt{\Lambda_p}(j+\frac{1}{2})}\n\\
&=&\sum_{k=0}^{\infty} (2k+1)e^{i\frac{n_O}{n}\sqrt{\Lambda}(2k+1)}\left(2\frac{\partial}{\partial(i\sqrt{\Lambda_p})}\sum_{j=|k-s|}^{k+s} e^{i\sqrt{\Lambda_p}(j+\frac{1}{2})}\right),
\ea
where we have introduced the renormalized cosmological constant $\sqrt{\Lambda_p}\equiv 2\frac{n_I}{n}\sqrt{\Lambda}$.

%\begin{equation}
%<s|\Psi_+>=\sum_{k=0}^{\infty} (2k+1)e^{i\sqrt{\Lambda}(k+\frac{1}{2})}
%\sum_{j=|k-s|}^{j=k+s}(2j+1)e^{i\sqrt{\Lambda}(j+\frac{1}{2})}
%=\sum_{k=0}^{\infty} (2k+1)e^{i\sqrt{\Lambda}(k+\frac{1}{2})}\left(2\frac{\partial}{\partial(i\sqrt{\Lambda})}\sum_{j=|k-s|}^{k+s} e^{i\sqrt{\Lambda}(j+\frac{1}{2})}\right)\,.
%\end{equation}
Let us now compute
$2\frac{\partial}{\partial(i\sqrt{\Lambda_p})}\sum_{j=|k-s|}^{k+s} e^{i\sqrt{\Lambda_p}(j+\frac{1}{2})}$. 
Assuming $k>s$\footnote{In the case $k<s$ we would obtain a finite sum but since then we need to normalize our state by diving for $<\emptyset |\Psi>$ this contribution would vanish since the normalization factor is divergent in $k$, as shown below.}, we have
\baa
2\frac{\partial}{\partial(i\sqrt{\Lambda_p})}\sum_{j=k-s}^{k+s} e^{i\sqrt{\Lambda_p}(j+\frac{1}{2})}
&=&2\frac{\partial}{\partial(i\sqrt{\Lambda_p})} e^{i\sqrt{\Lambda_p}(k+\frac{1}{2})}
\left(\frac{\sin(\sqrt{\Lambda_p}(s+\frac{1}{2}))}{\sin(\frac{\sqrt{\Lambda_p}}{2})}\right)\n\\
&=&(2k+1)e^{i\frac{n_I}{n}\sqrt{\Lambda}(2k+1)}
\left(\frac{\sin(\sqrt{\Lambda_p}(s+\frac{1}{2}))}{\sin(\frac{\sqrt{\Lambda_p}}{2})}\right)+
2e^{i\frac{n_I}{n}\sqrt{\Lambda}(2k+1)}
\frac{\partial}{\partial(i\sqrt{\Lambda_p})}
\frac{\sin(\sqrt{\Lambda_p}(s+\frac{1}{2}))}{\sin(\frac{\sqrt{\Lambda_p}}{2})}.
\eaa
%Let us now compute the case $k<l$:
%\begin{equation}
%2\frac{\partial}{\partial(i\sqrt{\Lambda})}\sum_{j=l-k}^{k+l} e^{i\sqrt{\Lambda}(j+\frac{1}{2})}=
%2\frac{\partial}{\partial(i\sqrt{\Lambda})} e^{i\sqrt{\Lambda}(l+\frac{1}{2})}
%\left(\frac{\sin(\sqrt{\Lambda}(k+\frac{1}{2}))}{\sin(\frac{\sqrt{\Lambda}}{2})}\right)=
%\end{equation}
%\begin{equation}
%=(2l+1)e^{i\sqrt{\Lambda}(l+\frac{1}{2})}
%\left(\frac{\sin(\sqrt{\Lambda}(k+\frac{1}{2}))}{\sin(\frac{\sqrt{\Lambda}}{2})}\right)
%+2e^{i\sqrt{\Lambda}(l+\frac{1}{2})}
%\frac{\partial}{\partial(i\sqrt{\Lambda})}
%\frac{\sin(\sqrt{\Lambda}(k+\frac{1}{2}))}{\sin(\frac{\sqrt{\Lambda}}{2})}=
%\end{equation}
%\begin{equation}
%=(2l+1)e^{i\sqrt{\Lambda}(l+\frac{1}{2})}
%\left(\frac{\sin(\sqrt{\Lambda}(k+\frac{1}{2}))}{\sin(\frac{\sqrt{\Lambda}}{2})}\right)+
%\end{equation}
%\begin{equation}
%+\left(\frac{e^{i\sqrt{\Lambda}(l+\frac{1}{2})}}{i}\right)\left((2k+1)
%\frac{\cos(\sqrt{\Lambda}(k+\frac{1}{2}))}{\sin(\frac{\sqrt{\Lambda}}{2})}
%-\frac{\sin(\sqrt{\Lambda}(k+\frac{1}{2}))\cos(\frac{\sqrt{\Lambda}}{2})}
%{\sin^2(\frac{\sqrt{\Lambda}}{2})}\right)
%\end{equation}
Therefore, the scalar product of the loop state with the first part of out state $\Psi$ gives
\ba\la{eq:con1}
<s|\Psi_+>&=&\frac{\sin(\sqrt{\Lambda_p}(s+\frac{1}{2}))}{\sin(\frac{\sqrt{\Lambda_p}}{2})}\sum_k (2k+1)^2e^{i\sqrt{\Lambda}(2k+1)}
+2\left(\frac{\partial}{\partial(i\sqrt{\Lambda_p})}
\frac{\sin(\sqrt{\Lambda_p}(s+\frac{1}{2}))}{\sin(\frac{\sqrt{\Lambda_p}}{2})}\right)\sum_k (2k+1)e^{i\sqrt{\Lambda}(2k+1)}.\n\\
\ea
%while for $k<l$ we have:
%\begin{equation}
%<\alpha_l|\Psi_+>=(2l+1)e^{i\sqrt{\Lambda}(l+\frac{1}{2})}
%\sum_{k=0}^{\infty} (2k+1)e^{i\sqrt{\Lambda}(k+\frac{1}{2})}
%\left(\frac{\sin(\sqrt{\Lambda}(k+\frac{1}{2}))}{\sin(\frac{\sqrt{\Lambda}}{2})}\right)+
%\end{equation}
%\begin{equation}
%+\left(\frac{e^{i\sqrt{\Lambda}(l+\frac{1}{2})}}{i}\right)
%\sum_{k=0}^{\infty} (2k+1)e^{i\sqrt{\Lambda}(k+\frac{1}{2})}
%\left((2k+1)\frac{\cos(\sqrt{\Lambda}(k+\frac{1}{2}))}{\sin(\frac{\sqrt{\Lambda}}{2})}
%-\frac{\sin(\sqrt{\Lambda}(k+\frac{1}{2}))\cos(\frac{\sqrt{\Lambda}}{2})}
%{\sin^2(\frac{\sqrt{\Lambda}}{2})}\right)\,,
%\end{equation}

An analogous calculation shows that for the second part of the state $\Psi$,
\bee
\Psi_-=\lim_{n\rightarrow \infty}\prod_p \sum_{j_p}(2j_p+1)\left(1-i\frac{\sqrt{\Lambda}}{n}(2j_p+1)\right)\chi_{j_p}(W_p), 
\eee
we obtain
\ba\la{eq:con2}
<s|\Psi_->&=&\frac{\sin(\sqrt{\Lambda_p}(s+\frac{1}{2}))}{\sin(\frac{\sqrt{\Lambda_p}}{2})}\sum_k (2k+1)^2e^{-i\sqrt{\Lambda}(2k+1)}
+2\left(\frac{\partial}{\partial(-i\sqrt{\Lambda_p})}
\frac{\sin(\sqrt{\Lambda_p}(s+\frac{1}{2}))}{\sin(\frac{\sqrt{\Lambda_p}}{2})}\right)\sum_k (2k+1)e^{-i\sqrt{\Lambda}(2k+1)}.\n\\
\ea
%while for $k<l$ we have:
%\begin{equation}
%<\alpha_l|\Psi_->=(2l+1)e^{-i\sqrt{\Lambda}(l+\frac{1}{2})}\sum_{k=0}^{\infty} (2k+1)e^{-i\sqrt{\Lambda}(k+\frac{1}{2})}
%\left(\frac{\sin(\sqrt{\Lambda}(k+\frac{1}{2}))}{\sin(\frac{\sqrt{\Lambda}}{2})}\right)+
%\end{equation}
%\begin{equation}
%+\left(\frac{e^{-i\sqrt{\Lambda}(l+\frac{1}{2})}}{-i}\right)
%\sum_{k=0}^{\infty} (2k+1)e^{-i\sqrt{\Lambda}(k+\frac{1}{2})}
%\left((2k+1)\frac{\cos(\sqrt{\Lambda}(k+\frac{1}{2}))}{\sin(\frac{\sqrt{\Lambda}}{2})}
%-\frac{\sin(\sqrt{\Lambda}(k+\frac{1}{2}))\cos(\frac{\sqrt{\Lambda}}{2})}{\sin^2(\frac{\sqrt{\Lambda}}{2})}\right)\,,
%\end{equation}
Hence, summing up the two contributions (\ref{eq:con1})-(\ref{eq:con2}), we get
\ba\la{eq:Loop-Psi}
<s|\Psi>&=&\frac{\sin(\sqrt{\Lambda_p}(s+\frac{1}{2}))}{\sin(\frac{\sqrt{\Lambda_p}}{2})}\sum_k (2k+1)^2 \cos\left(\sqrt{\Lambda}(2k+1)\right)\n\\
&+&2i\left(\frac{\partial}{\partial(\sqrt{\Lambda_p})}
\frac{\sin(\sqrt{\Lambda_p}(s+\frac{1}{2}))}{\sin(\frac{\sqrt{\Lambda_p}}{2})}\right)
\sum_k (2k+1) \sin\left(\sqrt{\Lambda}(2k+1)\right).
\ea
%while for the case $k<l$
%\begin{equation}
%<l|\Psi>=\frac{\sin(\sqrt{\Lambda}(l+\frac{1}{2}))}{\sin(\frac{\sqrt{\Lambda}}{2})}\sum_k (2k+1)^2\left(\cos\left(\sqrt{\Lambda}(k+\frac{1}{2})\right)\right)^{\frac{N}{n}}
%\cos\left(\sqrt{\Lambda}(k+\frac{1}{2})\right)+
%\end{equation}
%\begin{equation}
%+\left((l+\frac{1}{2})\frac{\cos(\sqrt{\Lambda}(l+\frac{1}{2}))}{\sin(\frac{\sqrt{\Lambda}}{2})}\right)
%\sum_k (2k+1)\left(\cos\left(\sqrt{\Lambda}(k+\frac{1}{2})\right)\right)^{\frac{N}{n}}
%\sin\left(\sqrt{\Lambda}(k+\frac{1}{2})\right)+
%\end{equation}
%\begin{equation}
%-\left(\frac{\cos(\frac{\sqrt{\Lambda}}{2})}
%{\sin(\frac{\sqrt{\Lambda}}{2})}
%\frac{\sin(\sqrt{\Lambda}(l+\frac{1}{2}))}{\sin(\frac{\sqrt{\Lambda}}{2})}\right)
%\sum_k (2k+1)\left(\cos\left(\sqrt{\Lambda}(k+\frac{1}{2})\right)\right)^{\frac{N}{n}}
%\sin\left(\sqrt{\Lambda}(k+\frac{1}{2})\right)
%\end{equation}

From the result (\ref{eq:Loop-Psi}) we see that the inner product of the state (\ref{eq:Ansatz}) with a loop state contains divergent constants. This feature is related to the topology of $\Sigma$ which we assumed to be a 2-sphere (Euler characteristic $\chi=2$) and it appears also in the case of vanishing cosmological constant. In fact, let us recall that in the $\Lambda=0$ case, when computing the physical scalar product between the vacuum and the one loop state, by means of the proper regularized version of the projector into the kernel of the curvature constraint, assuming the Cauchy surface representing space to have genus $g=0$, one gets \cite{Noui Perez}
\be
<s| \emptyset>_{ph}=<s|\prod_p\delta(U_p)\emptyset>=(2s+1)\sum_j(2j+1)^2,
\ee
where $U_p\in SU(2)$ is the holonomy around the plaquette $p$. The divergent constant is related to the redundancy in the product of delta distributions in the expression of the projection operator and is a consequence of the discrete analog of the Bianchi identity. The correct result can be obtained either by renormalizing the transition amplitude by the vacuum-vacuum physical inner product, $<\emptyset | \emptyset>_{ph}=\sum_j(2j+1)^2$, or by eliminating a single arbitrary plaquette holonomy $U_p$ from the product\footnote{Notice, for instance, that, in a similar fashion, the quantum dimension times a divergent constant is found in \cite{Sahlmann}, where the authors introduce a new, LQG inspired, technique to compute the expectation value of unknotted holonomy loops in $SU(2)$ Chern-Simons theory. The divergent constant is then eliminated by a normalization choice freedom.}. 

Here we proceed in an analogous way and we will see that the correct result is obtained renormalizing the scalar product $<s|\Psi>$ by $<\emptyset |\Psi>$, which, from the previous calculation, we can immediately compute, namely
\be\la{eq:Normalization}
<\emptyset |\Psi>=\sum_k (2k+1)^2 \cos\left(\sqrt{\Lambda}(2k+1)\right)\,.
\ee
Now, in order to deal with the rapport of divergent sums (since the $SU(2)$ $\Irrep$s are infinite dimensional), as those entering (\ref{eq:Loop-Psi}) and (\ref{eq:Normalization}),  and compute the normalized amplitude $<s|\Psi>/<\emptyset |\Psi>$, we will introduce a cut-off $k_M$ and then take the limit $k_M\rightarrow \infty$. If we do so, the final result for the scalar product between our ansatz (\ref{eq:Ansatz}) and a loop state $\alpha$ in the $\Irrep$ $s$ is
\ba\la{eq:Loop-Psi Ren}
\frac{<s|\Psi>}{<\emptyset |\Psi>}&=&\frac{\sin(\sqrt{\Lambda_p}(s+\frac{1}{2}))}{\sin(\frac{\sqrt{\Lambda_p}}{2})}\n\\
&+&2i\left(\frac{\partial}{\partial(\sqrt{\Lambda_p})}
\frac{\sin(\sqrt{\Lambda_p}(s+\frac{1}{2}))}{\sin(\frac{\sqrt{\Lambda_p}}{2})}\right)
\lim_{k_M\rightarrow \infty}\frac{\sum_{k=0}^{k_M} (2k+1)\sin\left(\sqrt{\Lambda}(2k+1)\right)}{ \sum_{k=0}^{k_M}(2k+1)^2 \cos\left(\sqrt{\Lambda}(2k+1)\right)}\n\\
&=&\frac{\sin(\sqrt{\Lambda_p}(s+\frac{1}{2}))}{\sin(\frac{\sqrt{\Lambda_p}}{2})}=[s]_q\,,
\ea
where $[s]_q$ is the {\it quantum dimension} in the $\Irrep$ $s$ and the second term on the r.h.s. of  (\ref{eq:Loop-Psi}) has been suppressed by the normalization factor. The transition amplitude (\ref{eq:Loop-Psi Ren}) is exactly what one obtains in the Turaev-Viro model, i.e. the quantum group modification to the classical amplitude $(2s+1)$ which one gets in the Ponzano-Regge model and in $2+1$ LQG without cosmological constant. The previous calculation suggests that the ansatz (\ref{eq:Ansatz}) has good chances to implement the right dynamics. Further evidence to this statement will be provided in the next section.

\begin{figure}[h!]
\centering
  \includegraphics[width=6cm,angle=360]{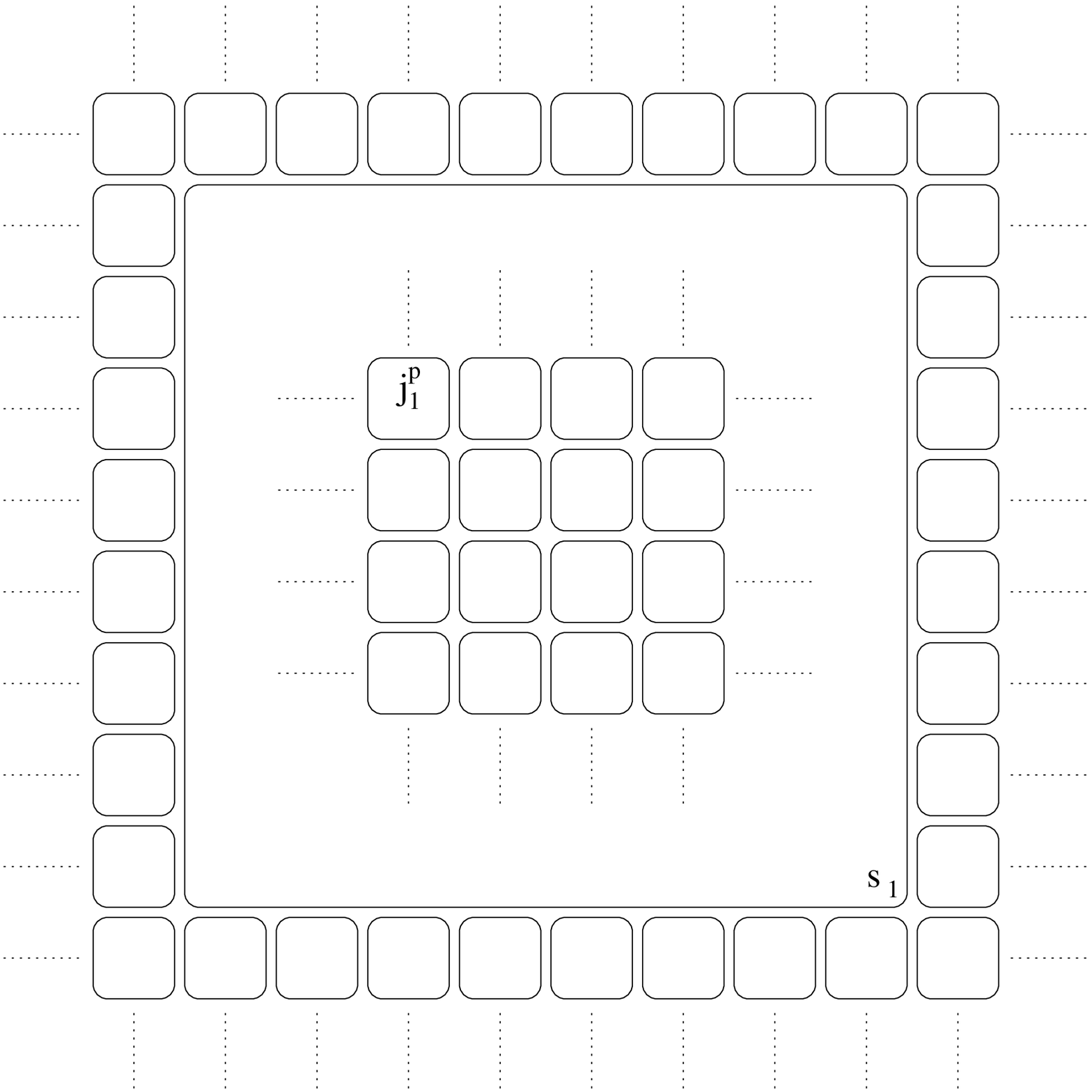} \includegraphics[width=6cm,angle=360]{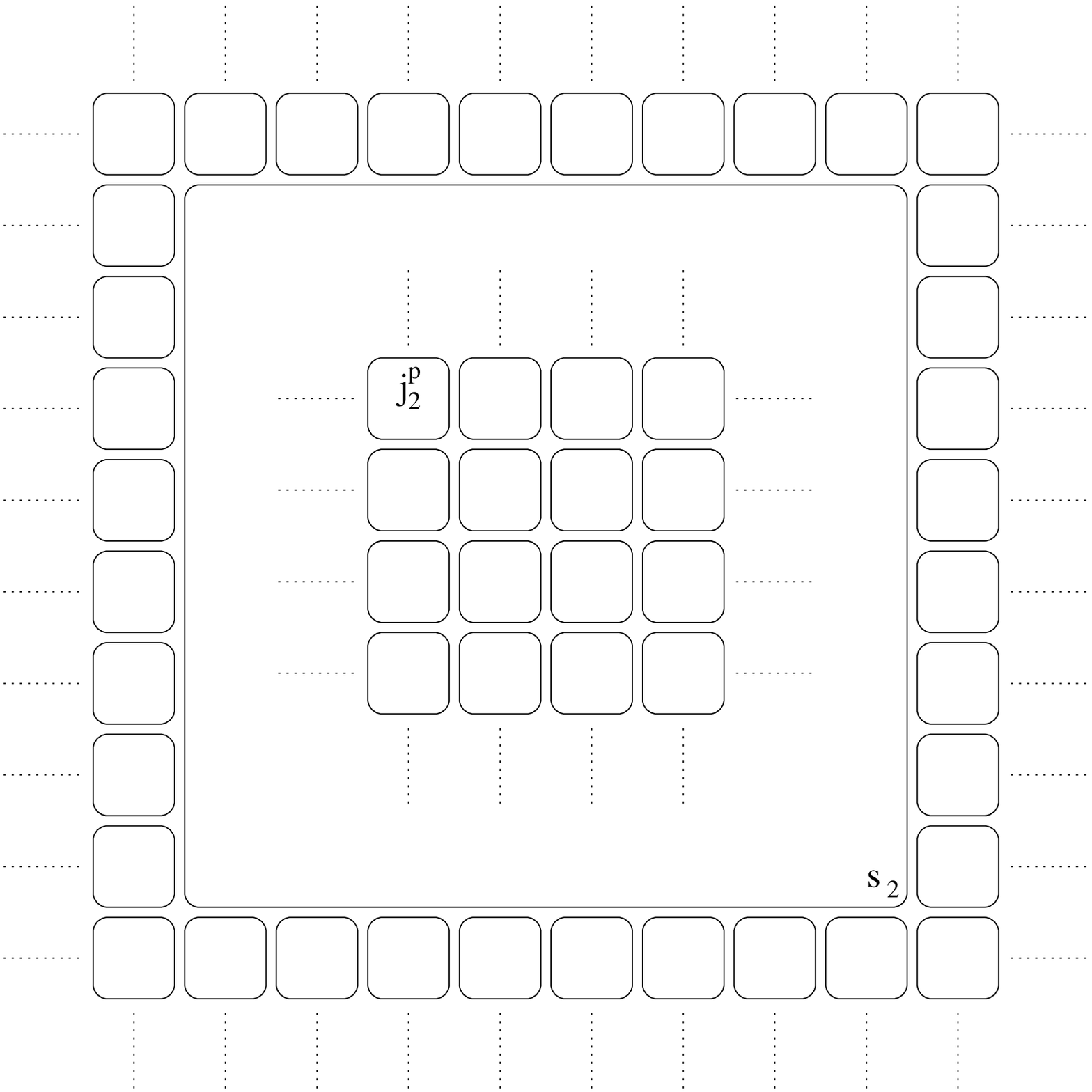}
  \caption{For a given $n$ there are finitely many $n_I$ plaquettes inside both loop states.}
  \label{sovra2}
\end{figure}

Let us now show that a very similar calculation to the one performed above gives, for the scalar product between the state (\ref{eq:Ansatz}) and multiple loops, the product of quantum dimensions in the irreducible representations associated to the loops.
We will show this for the case of a two loops state ($|s_1s_2>$); from the calculation it will be clear that the same result holds for a generic number of loops. As in the previous cases, we will concentrate first on the term $\Psi_+$ and then extend the calculation to the full state $\Psi$. In the calculation we will assume that the number of plaquettes inside the two loops is the same and we will denote this number by $n_I$, while the number of the remaining external plaquettes will be denoted by $n_O$; in this way, the renormalized cosmological constant enetering the definition of the deformation parameter $q$ for the two loops $s_1$ and $s_2$ will be the same. Given this prescription, we have
\ba
<s_1s_2|\Psi_+>&=&\int \left(\prod_h dg_h\right)\chi_{s_1}(g_\alpha)\chi_{s_2}(g_\beta) \lim_{n\rightarrow \infty}\prod_p \sum_{j_p}(2j_p+1)\left(1+i\frac{\sqrt{\Lambda}}{n}(2j_p+1)\right)\chi_{j_p}(W_p),
\ea
where $g_\alpha$ and $g_\beta$ are the holonomies around the teo loops $|s_1 s_2>$.
A graphical representation of the previous scalar product is given in FIG. (\ref{sovra2}).

As in the case of a single loop, we can now perform the integration over the group elements which are not associated to edges along the borders of the two loops. Once again, due to the relation (\ref{eq:integration}), all the plaquettes inside a given loop and all those outside both loops are forced to be in the same spin representation, which we denote, respectively, $j_1, j_2$ and $k$. Therefore, after this integrations we obtain
\baa
<\!s|\Psi_+\!>&=&\!\int\!\! \left(\prod_{\tilde{h}} dg_{\tilde{h}}\right)\!\!\lim_{n\rightarrow \infty}\sum_k \left(1+i\frac{\sqrt{\Lambda}}{n}(2k+1)\right)^{n_O}\sum_{j_1}(2j_1+1)\!\left(1+i\frac{\sqrt{\Lambda}}{n}(2j_1+1)\right)^{n_I}
\begin{array}{c}  \includegraphics[width=2cm,angle=360]{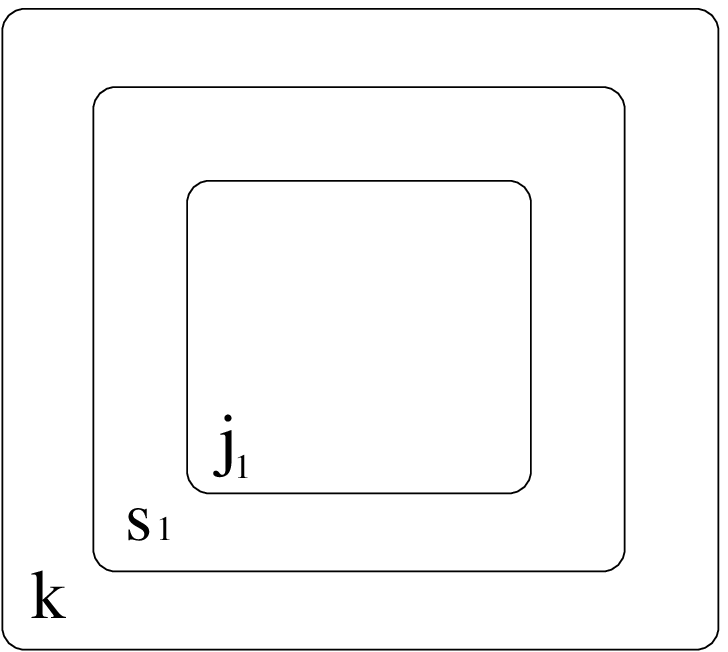}\end{array}\n\\
&\cdot&\sum_{j_2}(2j_2+1)\!\left(1+i\frac{\sqrt{\Lambda}}{n}(2j_2+1)\right)^{n_I}
\begin{array}{c}  \includegraphics[width=2cm,angle=360]{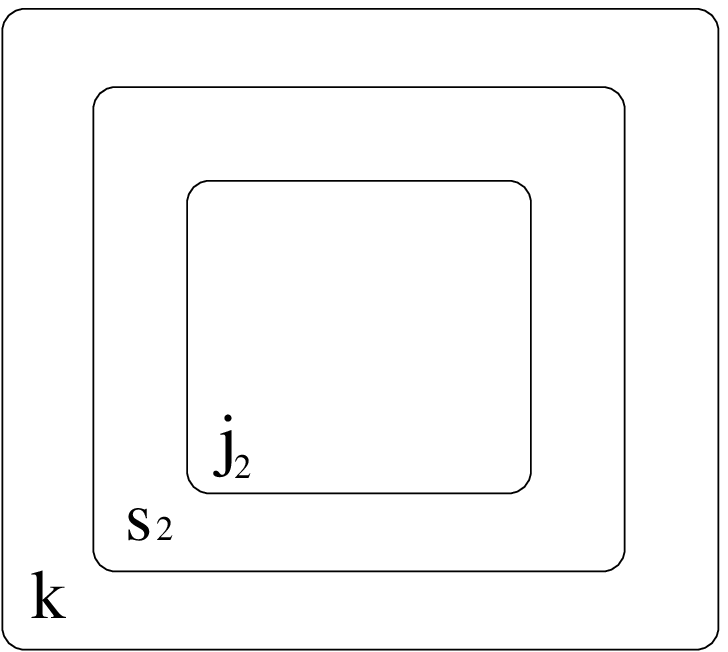}\end{array}\n\\
&=&\sum_{k=0}^{\infty} e^{i\frac{n_O}{n}\sqrt{\Lambda}(2k+1)}
\sum_{j_1=|k-s_1|}^{k+s_1}(2j_1+1)e^{i\sqrt{\Lambda_p}(j_1+\frac{1}{2})}
\sum_{j_2=|k-s_2|}^{k+s_2}(2j_2+1)e^{i\sqrt{\Lambda_p}(j_2+\frac{1}{2})}
\eaa
\ba
&=&\sum_{k=0}^{\infty} e^{i\frac{n_O}{n}\sqrt{\Lambda}(2k+1)}\left(2\frac{\partial}{\partial(i\sqrt{\Lambda_p})}\sum_{j_1=|k-s_1|}^{k+s_1} e^{i\sqrt{\Lambda_p}(j_1+\frac{1}{2})}\right)\left(2\frac{\partial}{\partial(i\sqrt{\Lambda_p})}\sum_{j_2=|k-s_2|}^{k+s_2} e^{i\sqrt{\Lambda_p}(j_2+\frac{1}{2})}\right)\n\\
&=&\frac{\sin(\sqrt{\Lambda_p}(s_1+\frac{1}{2}))}{\sin(\frac{\sqrt{\Lambda_p}}{2})}
\frac{\sin(\sqrt{\Lambda_p}(s_2+\frac{1}{2}))}{\sin(\frac{\sqrt{\Lambda_p}}{2})}
\sum_k (2k+1)^2e^{i\sqrt{\Lambda}(2k+1)}+ other~ terms,
\ea 
where again $\sqrt{\Lambda_p}=2\frac{n_I}{n}\sqrt{\Lambda}$ and the other terms in the scalar product are those suppressed by the normalization factor $<\emptyset |\Psi>$, as in the single loop case.
%, once, as in the single loop case, a cut-off for the label $k$ is introduced to do the sums and then the limit where the cut-off go to infinity is taken. 
Therefore, for the full normalized scalar product with a two-loop state we have
\be
\frac{<s_1 s_2|\Psi>}{<\emptyset |\Psi>}=\frac{\sin(\sqrt{\Lambda_p}(s_1+\frac{1}{2}))}{\sin(\frac{\sqrt{\Lambda_p}}{2})}
\frac{\sin(\sqrt{\Lambda_p}(s_2+\frac{1}{2}))}{\sin(\frac{\sqrt{\Lambda_p}}{2})}=[s_1]_q [s_2]_q.
\ee
\vskip0.5cm
{\it Remark}. In general, for the scalar product with a $\ell$-loops state the physical cosmological constant will scale in the same way, i.e. $\sqrt{\Lambda_p}=2\frac{n_I}{n}\sqrt{\Lambda}$, as long as the limit $n\rightarrow \infty$ is taken by having the same number of plaquettes $n_I$ inside each loop. Adopting this regularization scheme is necessary to have the deformation parameter $q$, entering the expression of the quantum dimension for a given Irrep, be the same for each loop in the final state. 

\section{Master constraint}\label{Master Constraint}

In order to prove that the state (\ref{eq:Ansatz}) introduced in the previous section implements the right dynamics, we are going to use the master constraint technique developed in \cite{Thiemann:2003zv} to deal with systems whose constraint algebra contains structure functions. More specifically,  
in this section we are going to show that, up to the first order in $\Lambda$, our ansatz (\ref{eq:Ansatz}) is a solution of the quantum version of the constraint:
\ba\la{eq:Master classic}
C^2&=&C^i C_i=(F^i(A)+\Lambda \epsilon^{ijk}E_j\wedge E_k)(F_i(A)+\Lambda \epsilon_i\!^{rs}E_r \wedge E_s)\n\\
&=&F^i(A)F_i(A)+ \Lambda \epsilon^{ijk}F_i(A)E_j \wedge E_k+\Lambda \epsilon^{ijk}E_j \wedge E_k F_i(A)+\Lambda^2(\epsilon^{ijk} E_j \wedge E_k)(\epsilon_i\!^{rs} E_r \wedge E_s)=0.
\ea
The previous constraint can be seen as a ``master constraint'' for $2+1$ gravity with $\Lambda\neq 0$. Since it is manifestaly gauge invariant, it commutes with the other constraint of the theory, namely the Gauss constraint $G_i\equiv D_a E^a_i=0$, and therefore we can try to impose strongly the quantum version of (\ref{eq:Master classic}) on spin network states, i.e states on which the Gauss constraint has already been imposed. 

With the decomposition of $\Sigma$ and the regularization scheme introduced in section \ref{Quantum Analysis}, we can now write the regularized  versions of the constraint (\ref{eq:Master classic}), namely
\begin{equation}\label{eq:Master discretized}
C^2_{\va R}=\sum_{p\in\CS} C^{ip}C_i^p,
\end{equation}
where $C^{ip}C_i^p$ is explicitly defined below.

The curvature-curvature term $F^i(A)F_i(A)$ is replaced by a 
Riemannian sum over all the placquettes of the square of the holonomy around the given placquette in the fundamental representation, namely:
\be\label{FF}
F^i(A)F_i(A)~~~~~~~~~\rightarrow~~~~~~~~~
\lim_{\varepsilon\rightarrow 0}\sum_p 4\frac{{\rm Tr}[W^p(A)\tau^i ]{\rm Tr}[W^p(A)\tau_i]}{\varepsilon^2},
\ee
where the contraction of the internal indices represents a link in the adjoint representation connecting the two holonomies around the same placquette $p$.

For the grasping terms inside the expression (\ref{eq:Master classic}) 
we write the regulated quantity corresponding to $\epsilon^{ijk}E_j E_k$ as
\be
\epsilon^{ijk}E^{\va R}_j \wedge E^{\va R}_k = \sum_{p} \epsilon^{ijk}(E_j^p \wedge E_k^p),
\ee
with
\ba
\epsilon^{ijk}(E_j^p \wedge E_k^p)&=&\Lambda \epsilon^{ijk}   \bigg( E_j(\eta_1)E_k(\eta_2)+  E_j(\eta_2)E_k(\eta_3)+
 E_j(\eta_3)E_k(\eta_4)+ E_j(\eta_4)E_k(\eta_1)\bigg)\n\\
&=&\Lambda \epsilon^{ijk}   \left( \sum_{a=1}^4 E_j(\eta_a)E_k(\eta_{a+1})\right),
\ea
where $\eta_{i}\in \CS^{\va *}$ are the four shifted edges shown in FIG.  \ref{fig:Cellular_decomposition} that are dual to the shadowed plaquette
$p\in \CS$ and in the sum, for the index $a$, we used the convention $a+1=5$ corresponds to $a=1$.

With this prescription, we can now regularize the remaining terms inside (\ref{eq:Master classic}), namely
\ba
&&\Lambda \epsilon^{ijk}F_i(A)E_j \wedge E_k~~~~~~~~~~~~~~~~~~~~~~~\rightarrow~~~~~~~~~~\lim_{\varepsilon\rightarrow 0}\frac{\Lambda}{\varepsilon^2}\sum_p
-{\rm Tr}[W^p(A)\tau_i ] \left( \sum_{a=1}^4 \epsilon^{ijk}E_j(\eta_a)E_k(\eta_{a+1}) \right)
\n\\
\n\\
\n\\
&&\Lambda \epsilon^{ijk}E_j \wedge E_k F_i(A)~~~~~~~~~~~~~~~~~~~~~~~\rightarrow~~~~~~~~~~\lim_{\varepsilon\rightarrow 0}\frac{\Lambda}{\varepsilon^2}\sum_p
-\left( \sum_{a=1}^4 \epsilon^{ijk}E_j(\eta_a)E_k(\eta_{a+1})\right){\rm Tr}[W^p(A)\tau_i ]
\n\\
\n\\
\n\\
&&\Lambda^2(\epsilon^{ijk} E_j \wedge E_k)(\epsilon_i\!^{rs} E_r \wedge E_s)~~~~~~~~~~\rightarrow~~~~~~~~~~\lim_{\varepsilon\rightarrow 0} \frac{\Lambda^2}{\varepsilon^2}\sum_p
\frac{\left( \sum_{a=1}^4 \epsilon^{ijk} E_j(\eta_a)E_k(\eta_{a+1})\right)
\left( \sum_{b=1}^4 \epsilon_i\!^{rs}E_r(\eta_b)E_s(\eta_{b+1})\right)}{4}.\n\\\label{eq:EEEE}
\ea
The quantities (\ref{FF}), (\ref{eq:EEEE}), entering the definition of the regularized version of the master constraint (\ref{eq:Master classic}), all have the right naive continuum limit.

The quantization of (\ref{eq:Master discretized}) then follows by promoting holonomies and electric fields to operators according to (\ref{ggcc})-(\ref{fluxx}). 

We are now ready to show that our state (\ref{eq:Ansatz}) is annihilated, up to the first order in $\Lambda$, by the quantum master constraint $\hat{C}^2$. 
From the expression (\ref{eq:Ansatz}) of the state $|\Psi>$ it is immediate to see that it contains only integer orders in $\Lambda$ terms, there are, therefore, only three contributions to the first order in $\Lambda$ of $\hat{C}^2 \triangleright |\Psi>$, explicitly
\ba\label{eq:First order}
(\hat{C}^2 \triangleright |\Psi>)^{(1)}&=&\lim_{\varepsilon\rightarrow 0}\sum_p\Bigg( 4\frac{{\rm Tr}[W^p(A)\tau^i ]{\rm Tr}[W^p(A)\tau_i]}{\varepsilon^2} \triangleright |\Psi^{(1)}>
+\n\\
&-&\frac{\Lambda}{\varepsilon^2}\sum_p
{\rm Tr}[W^p(A)\tau_i ] \left( \sum_{a=1}^4 \epsilon^{ijk}E_j(\eta_a)E_k(\eta_{a+1}) \right)
\triangleright|\Psi^{(0)}>\n\\
&-& \frac{\Lambda}{\varepsilon^2}\sum_p
\left( \sum_{a=1}^4 \epsilon^{ijk}E_j(\eta_a)E_k(\eta_{a+1})\right){\rm Tr}[W^p(A)\tau_i ]\triangleright|\Psi^{(0)}>\Bigg),
\ea
where $(m)$, with $m$ integer, indicates the component of $m{\rm th}$ order in $\Lambda$. In order to prove  $(\hat{C}^2 \triangleright |\Psi>)^{(1)}=0$, we are going to show that 
\be\label{eq:Generic first order}
<\phi|(\hat{C}^2 \triangleright |\Psi>)^{(1)}=0~~~~~~\forall \phi\in \Ha,
\ee
where the states $\phi\in \Ha$ will be a subset $Cyl (\CS)\subset Cyl$
consisting of all cylindrical functions whose underlying graph is built on the same discrete structure used to regularize the master constraint.

To prove that the scalar product (\ref{eq:Generic first order}) vanishes, it is enough to prove that it does so for each single placquette inside the sum over $p$ in the definition of $\hat{C}^2$. Before considering the scalar product with a generic state $\phi\in \Ha$, let us consider the case in which $\phi$ is a single Wilson loop around a placquette $p$ in the $\Irrep$ $s$. 
%We will see that this is enough to prove (\ref{eq:Generic first order}) for the generic case too. 
Moreover, we will first consider the action of the master constraint on the component $\Psi_+$ of $\Psi$, a completely analogous calculation follows for $\Psi_-$. In other words, what we are now going to compute is
\ba\label{eq:First order loop}
<s|\Bigg[&&4{\rm Tr}[W^p(A)\tau^i ]{\rm Tr}[W^p(A)\tau_i] \triangleright |\Psi_+^{(1)}>
\n\\
&-&\Lambda\Bigg(
{\rm Tr}[W^p(A)\tau_i ] \left( \sum_{a=1}^4 \epsilon^{ijk}E_j(\eta_a)E_k(\eta_{a+1}) \right)
+\left( \sum_{a=1}^4 \epsilon^{ijk}E_j(\eta_a)E_k(\eta_{a+1})\right){\rm Tr}[W^p(A)\tau_i ]\Bigg) \triangleright|\Psi_+^{(0)}>\Bigg]\n\\
&=&<s|\left[ \hat{C}^{(0)}\triangleright |\Psi_+^{(1)}>+\hat{C}^{(1)}\triangleright |\Psi_+^{(0)}>\right],
\ea
where $|s>$ represents a loop state in the $\Irrep$ $s$ around the placquette $p$ and
\ba
\hat{C}^{(0)}&=&4{\rm Tr}[W^p(A)\tau^i ]{\rm Tr}[W^p(A)\tau_i]\label{eq:C0}\\
\hat{C}^{(1)}&=&-\Lambda\,
{\rm Tr}[W^p(A)\tau_i ] \left( \sum_{a=1}^4 \epsilon^{ijk}E_j(\eta_a)E_k(\eta_{a+1}) \right) + \Lambda \left( \sum_{a=1}^4 \epsilon^{ijk}E_j(\eta_a)E_k(\eta_{a+1})\right){\rm Tr}[W^p(A)\tau_i ]\label{eq:C1}\\
\hat{C}^{(2)}&=&\frac{\Lambda^2}{4}
\left( \sum_{a=1}^4 \epsilon^{ijk} E_j(\eta_a)E_k(\eta_{a+1})\right)
\left( \sum_{b=1}^4 \epsilon_i\!^{rs}E_r(\eta_b)E_s(\eta_{b+1})\right).\label{eq:C2}
\ea
At this point, we will assume that the discrete structure on which the state $\Psi$ is built is much more refined that the one used to regularize the master constraint. This means that inside the placquette $p$ on which we have defined the quantum regularized version of the master constraint there are many of the loops that form our state $\Psi$. That's why, from now on, we will denote by $\tilde{p}$ the plaquettes on which the state $\Psi$ is defined and by $p$ those on which the master constraint $\hat{C}^2$ is regularized. This is justified by the fact that the limit $\tilde{n}\rightarrow\infty$ (where we have used the tilde over $n$ to be consistent with the notation just introduced) inside the expression (\ref{eq:Ansatz}) of the state $\Psi$ has to be taken before the limit $\varepsilon\rightarrow 0$ in the expression for the quantum master constraint $\hat{C}^2$. Besides, if one reversed the order of these two limits, he would get that the relation $<\phi|(\hat{C}^2 \triangleright |\Psi>)^{(1)}=0$ would be trivially satisfied. Another way of viewing this is that we are interested in imposing the master constraint not on a single component (loop) of $\Psi$ but on a ``chunk'' of it. This can be motivated by the fact that, in the limit $\tilde{n}\rightarrow \infty$, on a single loop the state looks like a flat state and in fact the quantum version of the flat curvature constraint $F^i(A)=0$ on a single loop component of $\Psi$ is immediately satisfied.

Before starting to compute $<s|\hat{C}^{(0)} \triangleright |\Psi_+^{(1)}>$, let us notice that, in order to select the first order in $\Lambda$ on $\Psi_+$, we need to pick from two of the many loops that form the state the contribution $i\frac{\sqrt{\Lambda}}{\tilde{n}}(2j_{\tilde{p}}+1)$. Now, there are three possibilities to make this choice: we can pick these contributions from two loops both outside the placquette $p$, both inside or one inside and one outside. We will consider only the last of these three possibilitites since it is the only one which will give a scalar product different from zero, as it will be clearer in a few steps. Let us start the computation: 

\ba
<s| \hat{C}^{(0)}  \triangleright |\Psi_+^{(1)}>&=&4\int \left(\prod_h dg_h\right)\chi_s(g_\alpha)  {\rm Tr}[W^p(A)\tau^i ]{\rm Tr}[W^p(A)\tau_i]  \triangleright\n\\
&&\left(\lim_{\tilde{n}\rightarrow \infty}\prod_{\tilde{p}} \sum_{j_{\tilde{p}}}(2j_{\tilde{p}}+1)\left(1+i\frac{\sqrt{\Lambda}}{\tilde{n}}(2j_{\tilde{p}}+1)\right)\chi_{j_{\tilde{p}}}(W_{\tilde{p}})\right)^{(1)}
\n\\
&=&4\int \left(\prod_{\tilde{h}} dg_{\tilde{h}}\right)\Bigg(\lim_{\tilde{n}\rightarrow \infty}\sum_{j,k}(2j+1)(2k+1)
\left(-\frac{\Lambda}{\tilde{n}^2}\frac{\tilde{n}}{2}\frac{\tilde{n}}{2}\right)
(2j+1)(2k+1)\Bigg) \begin{array}{c}  \includegraphics[width=3cm,angle=360]{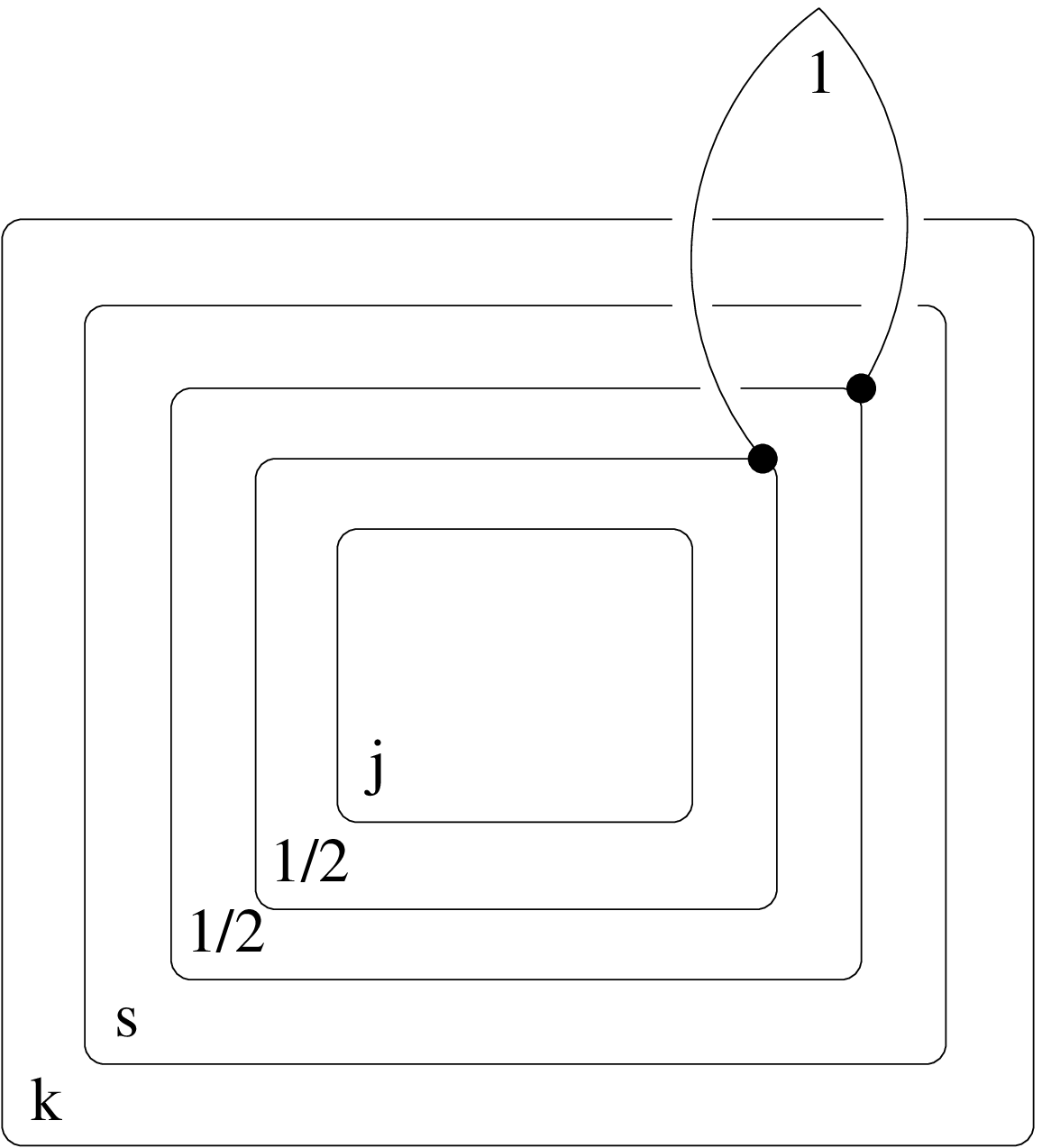}\end{array},\n\\
\ea
where we have used again the relation (\ref{eq:integration}) for the integration over the group elements associated to the internal and external edges which are not along the border of $p$ and we have called $k$ and $j$ the spin which, respectively, all the loops outside  and inside $p$ have in common due to the integration. The factor $\tilde{n}/2\cdot \tilde{n}/2$ comes from all the possible ways to pick up the two contributions that give the first order in $\Lambda$, where we have assumed that the number of loops inside and outside $p$ is the same.

If we now perform the integral over the remaining group elements and take the limit, we get:
\ba
<s| \hat{C}^{(0)}  \triangleright |\Psi_+^{(1)}>&=&
-\Lambda\sum_{j,k}(2j+1)^2(2k+1)^2
\begin{array}{c}\includegraphics[width=2.8cm,angle=360]{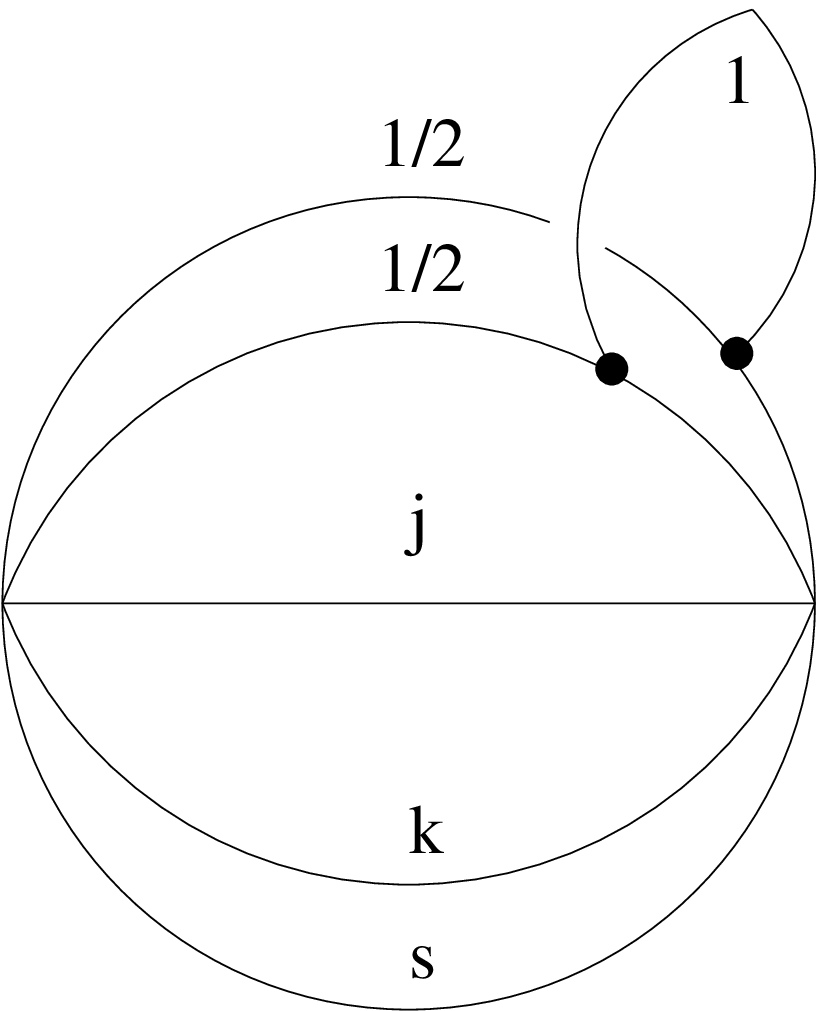}\end{array}.\n
\ea
We can now see how the two contributions coming from the choice of both loops (contributing to the first order in $\Lambda$) inside or outside $p$ would have given a zero scalar product. In this two cases, in fact, we can use the semi-semplicity relation:
\be\la{eq:Semisemplicity}
\sum_i (2i+1)\begin{array}{c}  \includegraphics[width=1.2cm,angle=360]{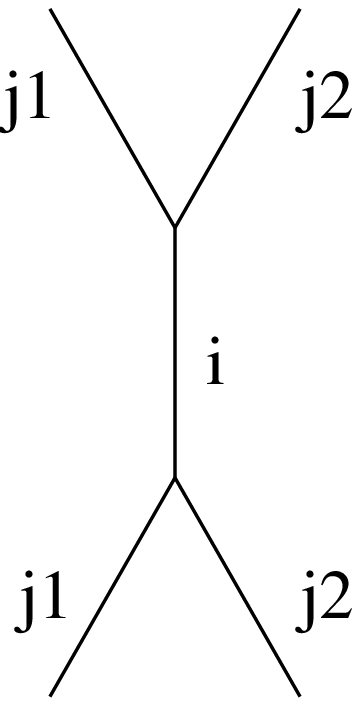}\end{array}=
\begin{array}{c}  \includegraphics[width=1.8cm,angle=360]{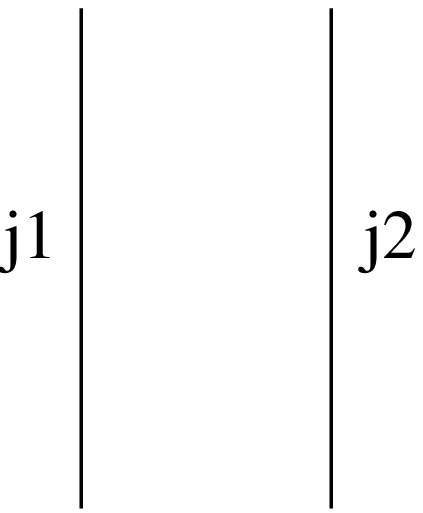}\end{array}
\ee

\noindent to get a spin network which evaluates to zero, namely of the form
\be\label{fig:box1box}
\begin{array}{c}  \includegraphics[width=5cm,angle=360]{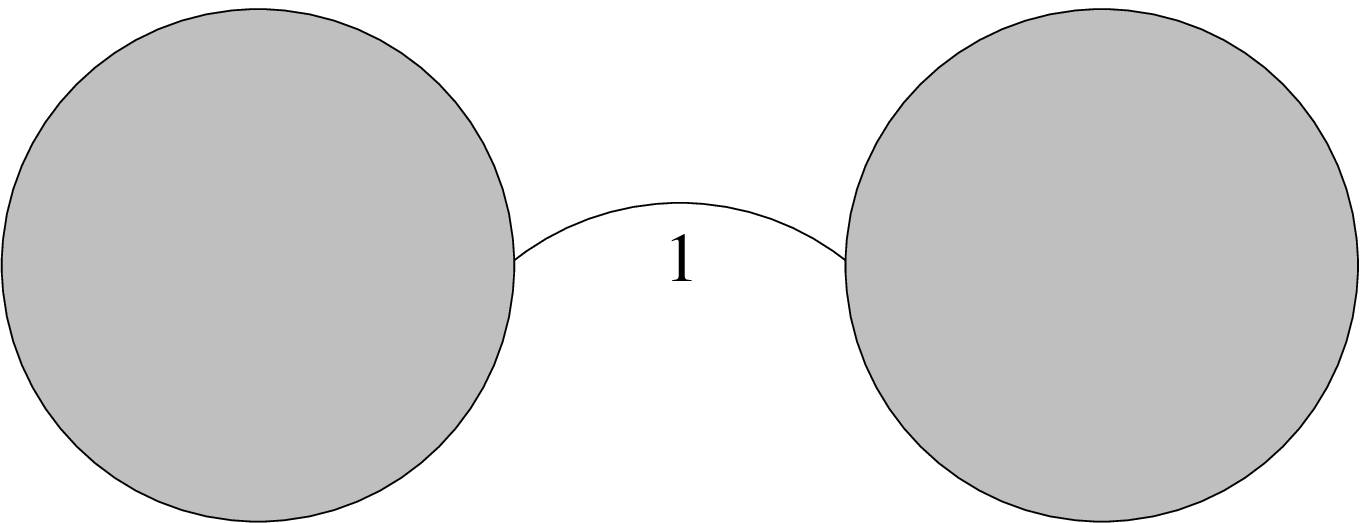}\end{array},
\ee
where the filled circles represent generic spin networks. 

To continue the calculation, let us introduce the following useful relations
\be
\begin{array}{c}  \includegraphics[width=3.5cm,angle=360]{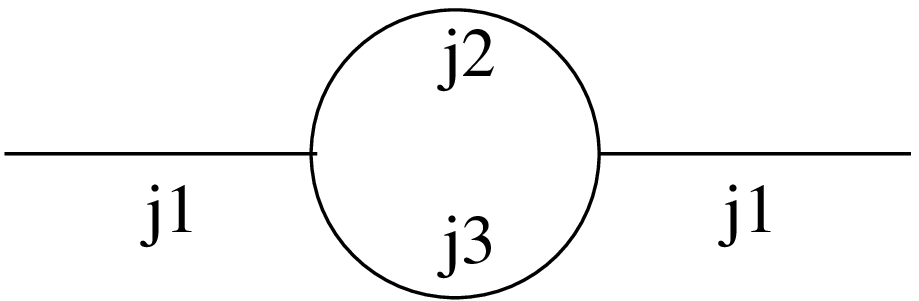}\end{array}
=\frac{1}{2j_1+1}  \begin{array}{c} \includegraphics[width=2.5cm,angle=360]{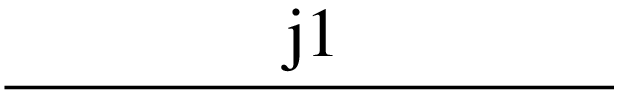}\end{array},
\ee
\be
\begin{array}{c}  \includegraphics[width=2cm,angle=360]{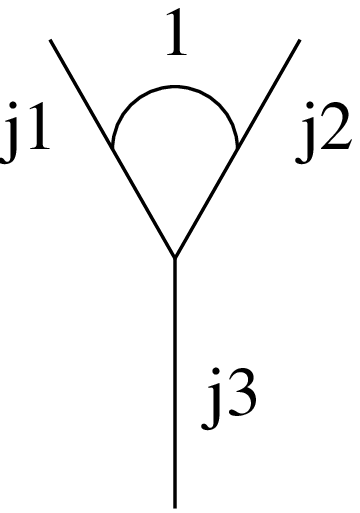}\end{array}
=-\frac{1}{2}(j_3(j_3+1)-j_1(j_1+1)-j_2(j_2+1))  \begin{array}{c} \includegraphics[width=2cm,angle=360]{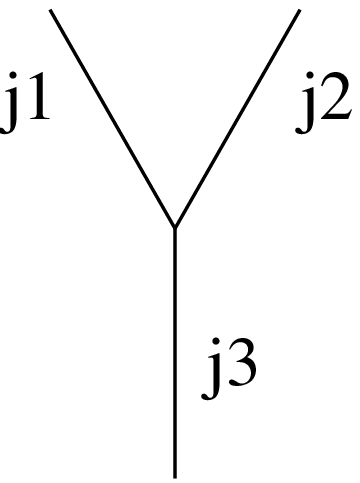}\end{array},
\ee
and
\be
\begin{array}{c}  \includegraphics[width=3.5cm,angle=360]{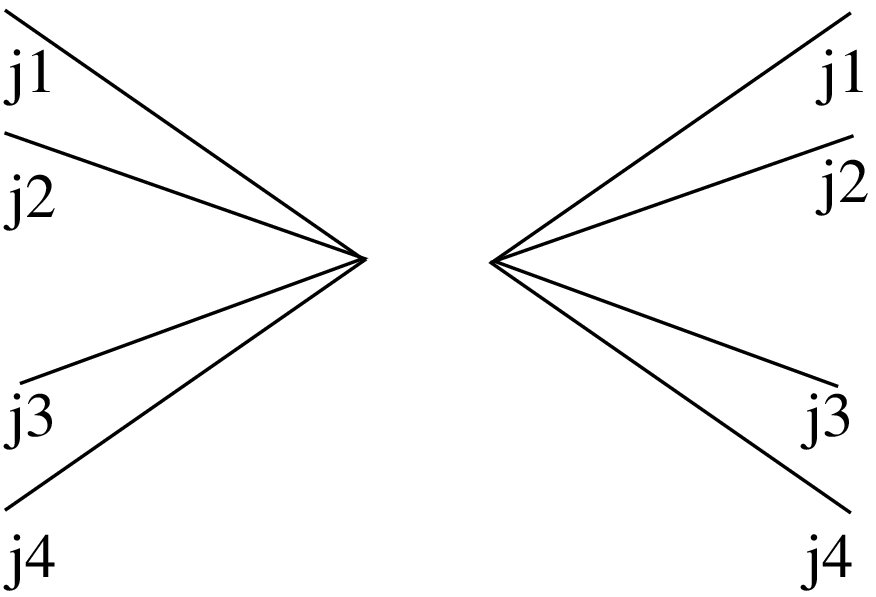}\end{array}
=\sum_i (2i+1)\begin{array}{c}  \includegraphics[width=3.5cm,angle=360]{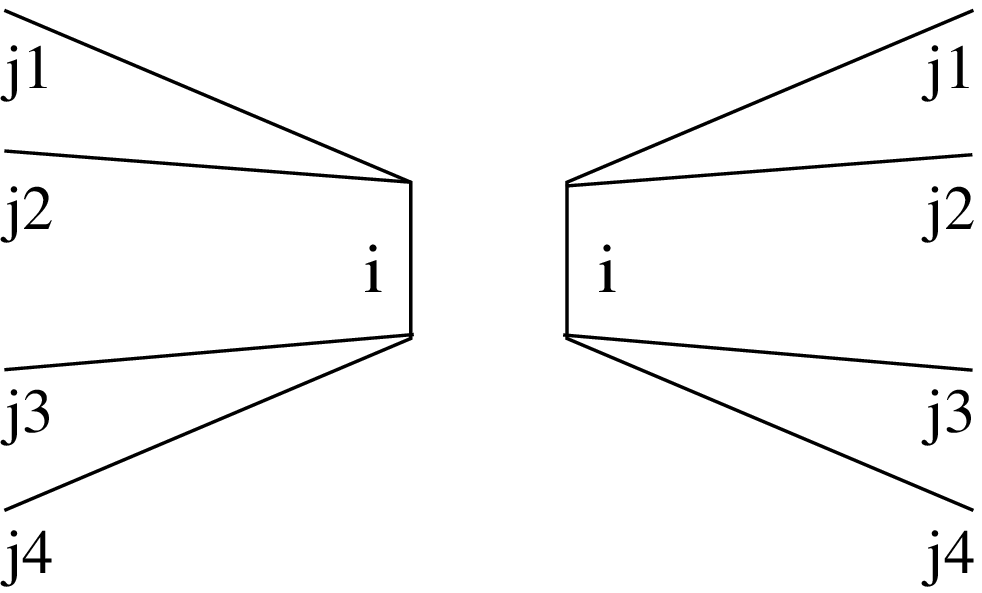}\end{array};
\ee
 By means of them, we have:
\ba
<s|  \hat{C}^{(0)} \triangleright |\Psi_+^{(1)}>&=&
\frac{\Lambda}{2}\sum_{j,k}(2j+1)^2(2k+1)^2\sum_i (2i+1)\left(i(i+1)-\frac{3}{2})\right)
\begin{array}{c}  \includegraphics[width=2.5cm,angle=360]{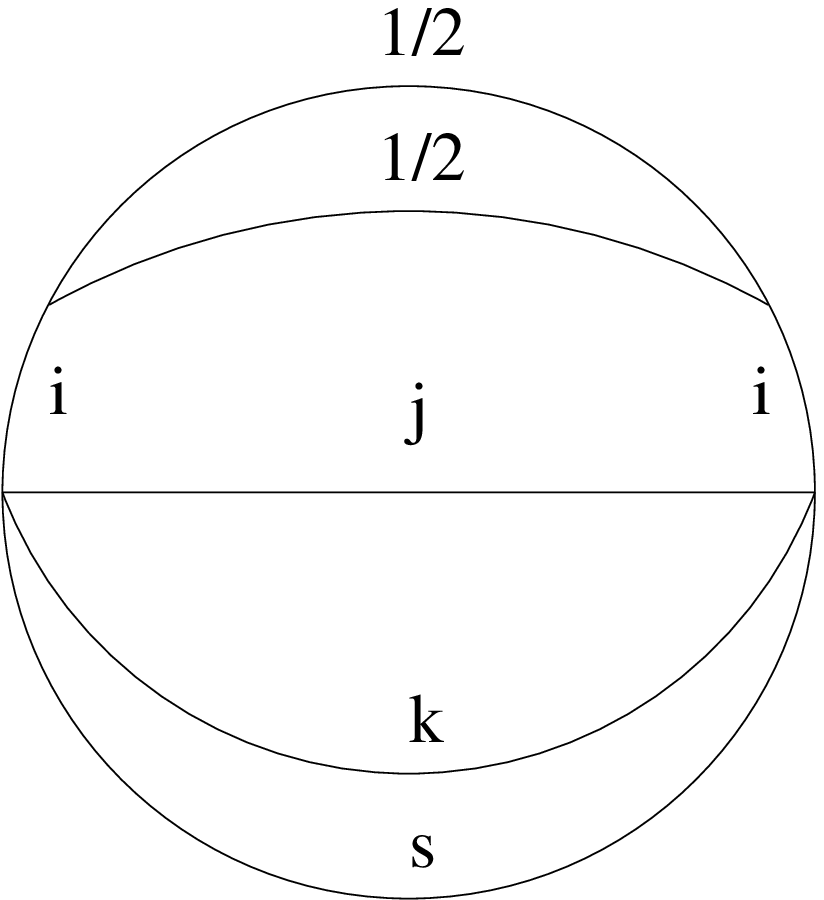}\end{array}\n\\
&=&\frac{\Lambda}{2}\sum_{j,k}(2j+1)^2(2k+1)^2\Bigg(\frac{1}{2}\begin{array}{c}  \includegraphics[width=2.2cm,angle=360]{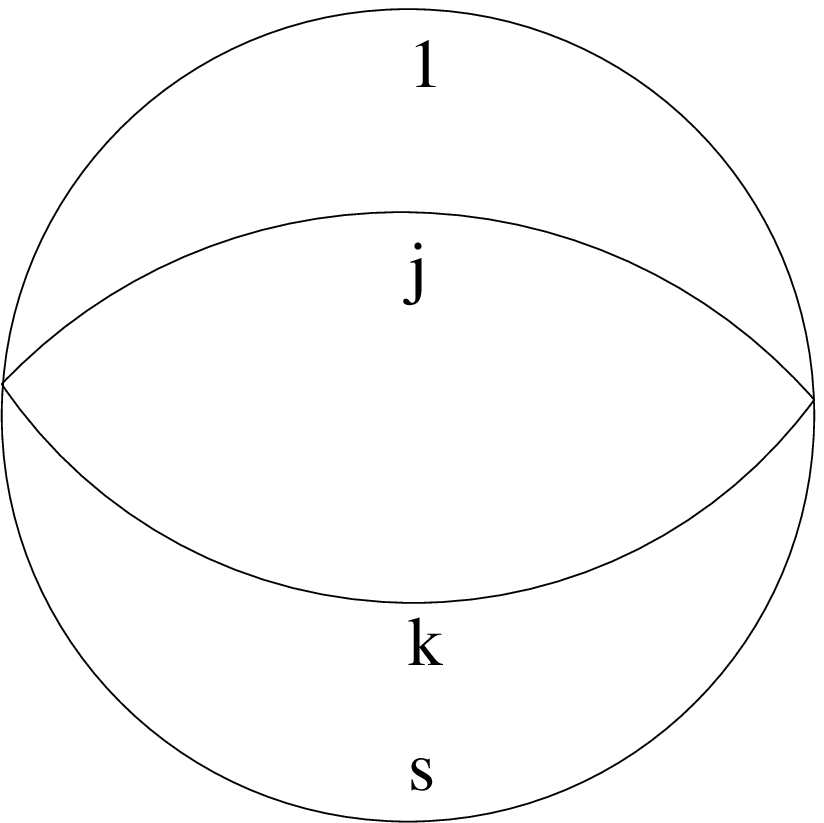}\end{array}
-\frac{3}{2}\begin{array}{c}  \includegraphics[width=2.2cm,angle=360]{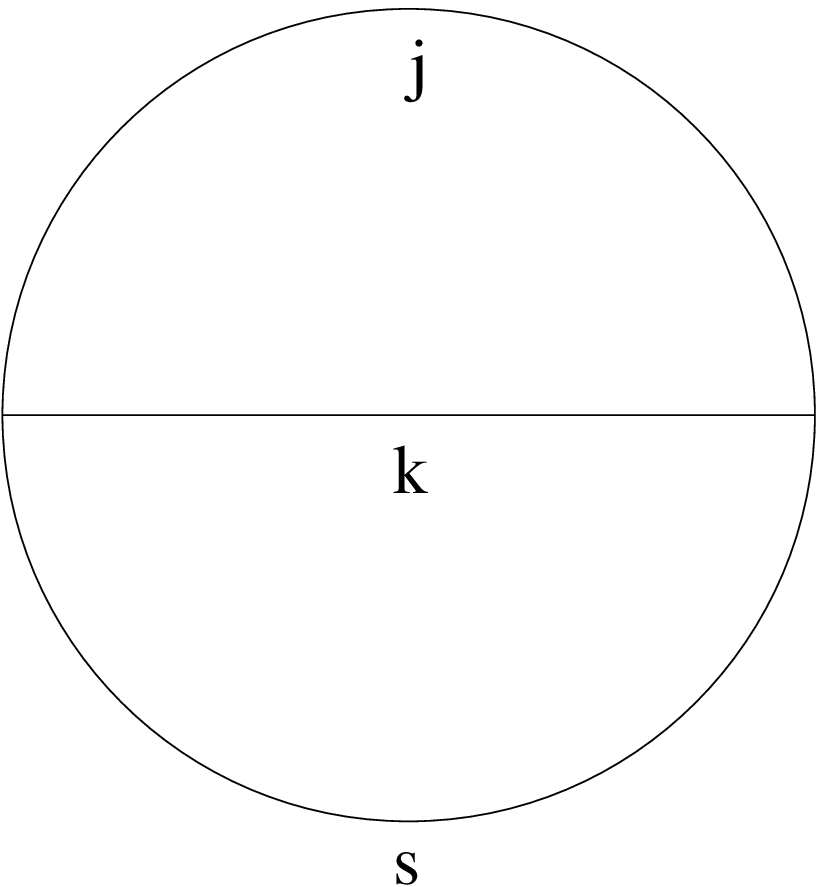}\end{array}\Bigg)\n\\
&=&2\Lambda\sum_j(2j+1)^2(2s+1).
\ea
Let us now compute the other two contributions to the scalar product (\ref{eq:First order loop}). In this case the two operators  entering the definition of the master constraint act on the zeroth order in $\Lambda$ component of $\Psi_+$ and this correspond to the product of delta functions on the group $SU(2)$ one for each placquette, i.e. on the flat state. This simplifies a lot the calculation of these other two contributions. In fact, all scalar products between the loop $|s>$ and the states obtained by the action of the first term in $\hat{C}^{(1)}$ (see (\ref{eq:C1})) on $|\Psi_+^{(0)}>$ give rise to spin networks of the form pictured in (\ref{fig:box1box}) and therefore give zero contributions. The same happens for the second term in $\hat{C}^{(1)}$  except for the fact that now there is one more state created by its action on $|\Psi_+^{(0)}>$ and this gives a contribution different from zero. This state is the one obtained when the graspings of the two electric fields are both on the new placquette created by the previous action of the Wilson loop operator on the right. So, let us compute explicitly this contribution:
\ba
<s| \hat{C}^{(1)}\triangleright |\Psi_+^{(0)}>&=& -\Lambda\int \left(\prod_h dg_h\right)\chi_s(g_\alpha)\left( \sum_{a=1}^4 \epsilon^{ijk}E_j(\eta_a)E_k(\eta_{a+1})\right){\rm Tr}[W^p(A)\tau_i ]\triangleright\n\\
 &&\left(\lim_{\tilde{n}\rightarrow \infty}\prod_{\tilde{p}} \sum_{j_{\tilde{p}}}(2j_{\tilde{p}}+1)\left(1+i\frac{\sqrt{\Lambda}}{\tilde{n}}(2j_{\tilde{p}}+1)\right)\chi_{j_{\tilde{p}}}(W_{\tilde{p}})\right)^{(0)}\n\\
 &=& 4\Lambda\int \left(\prod_{\tilde{h}} dg_{\tilde{h}}\right)\sum_{j,k}(2j+1)(2k+1)
 \begin{array}{c}  \includegraphics[width=3cm,angle=360]{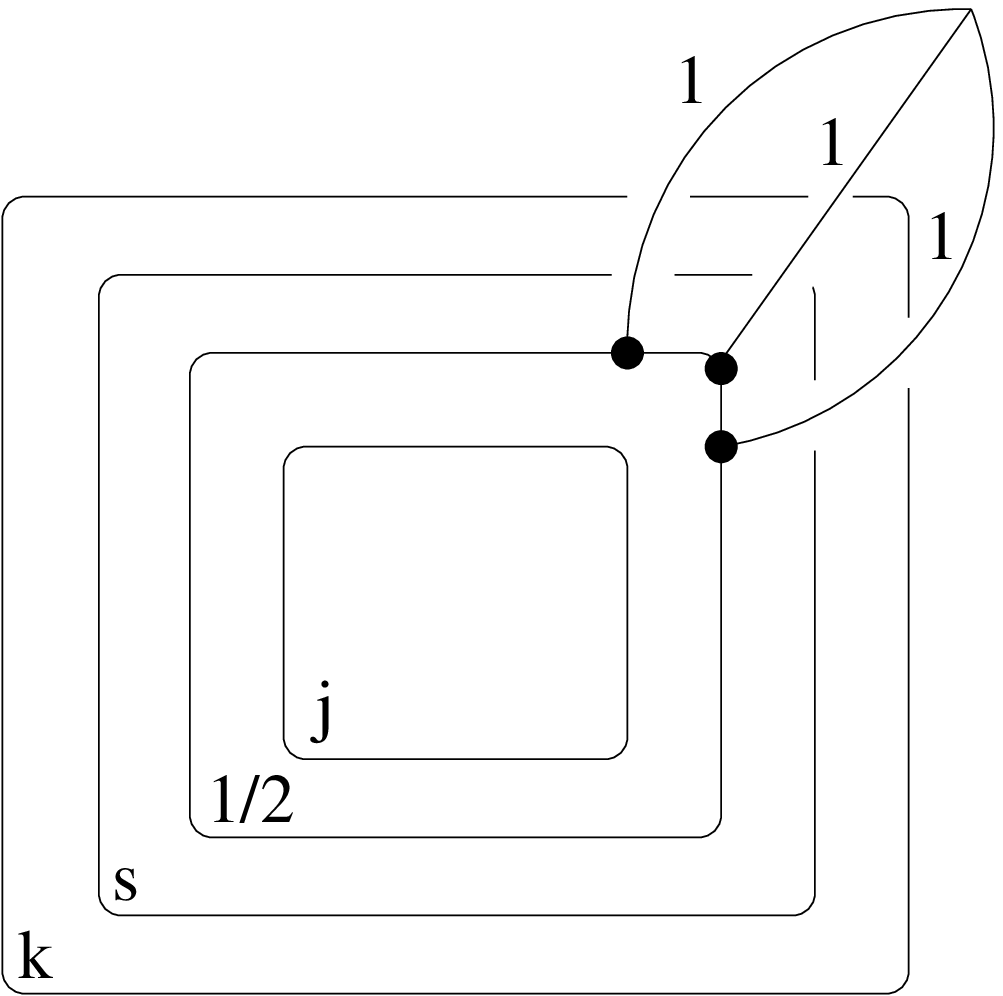}\end{array}\n\\
&=& -\Lambda \sum_{j,k}(2j+1)(2k+1)\sum_i(2i+1)\begin{array}{c}\includegraphics[width=2.6cm,angle=360]{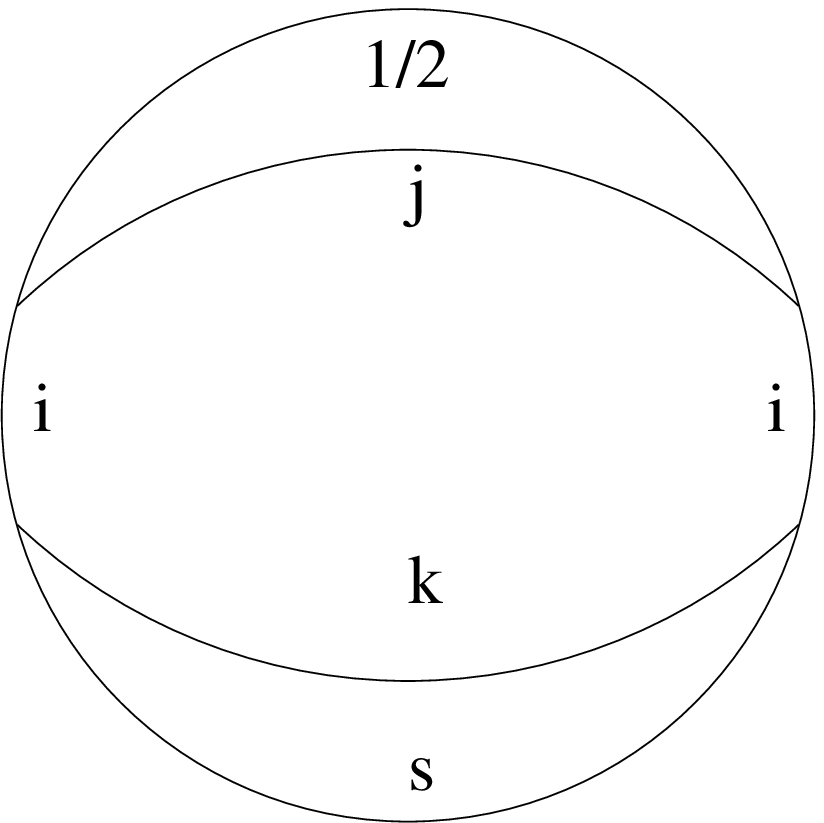}\end{array}\n\\
&=& -2\Lambda\sum_j(2j+1)^2(2s+1),
\ea
where in the last passage we have used twice the semi-semplicity relation (\ref{eq:Semisemplicity}) and in the third line we have adopted the same symmetrization scheme introduced in \cite{Anomaly}, namely summing over the three possible position of the link between the loop $W_i^p$ and the two graspings $E_j^p, E_k^p$ and dividing by three.

Therefore we see that the only non-vanishing contribution coming from the terms of the master constraint proportional to $\Lambda$ is exactly of the same form as the one coming from the term which has non graspings and the two cancel each others, proving the desired result
\be
<s|(\hat{C}^2 \triangleright |\Psi>)^{(1)}=0.
\ee

\begin{figure}[h!]
\centering
  \includegraphics[width=5cm,angle=360]{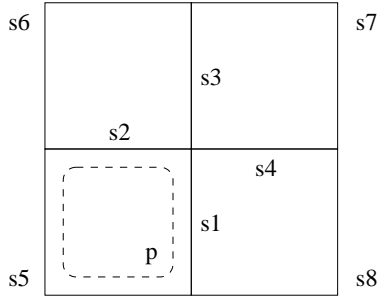}
  \caption{Generic state $|\phi>$ formed by 4 loops.}
  \label{fig:generic state}
\end{figure}

To show that the scalar product of $(\hat{C}^2 \triangleright |\Psi>)^{(1)}$ with a generic state $|\phi>$ also vanishes, let us consider the case in which we add three more loops around $<s|$, namely the state $|\phi>$ now has the form shown in FIG. \ref{fig:generic state}, where $s_1\cdots s_8$ are the spins associated to the different edges of the state and $p$ is the plaquette on which we are considering the constraint.
\\

Let us now show how, taking the scalar product with this more generic state, one obtains, for the two different non-vanishing contributions, the same result as for the single loop state times a spin network which can be factorized out and therefore doesn't change the result. In order to do so, we are going to use the following recoupling theory relations to simplify spin network evaluations:
\be
\begin{array}{c}\includegraphics[width=10cm,angle=360]{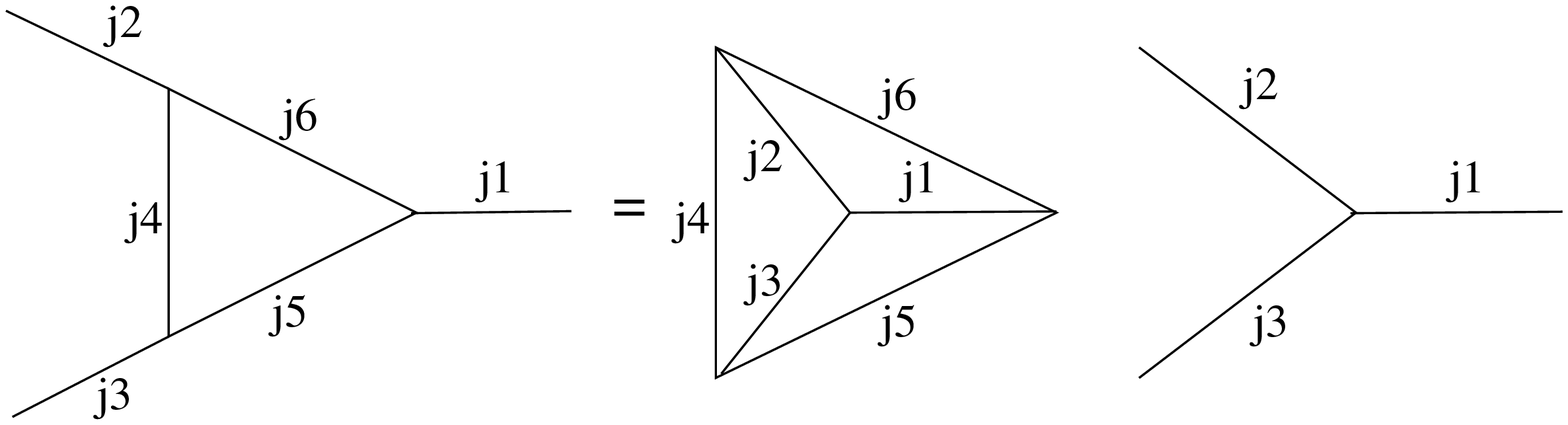}\end{array}
\ee
\be
\begin{array}{c}\includegraphics[width=3.5cm,angle=360]{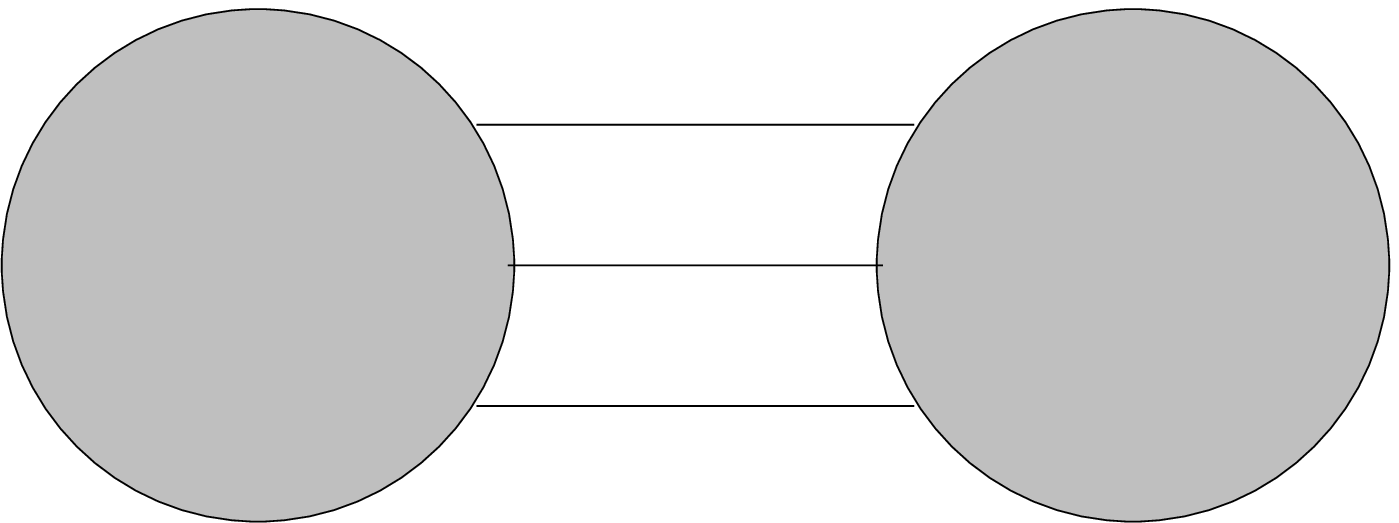}\end{array}=
\begin{array}{c}\includegraphics[width=3.5cm,angle=360]{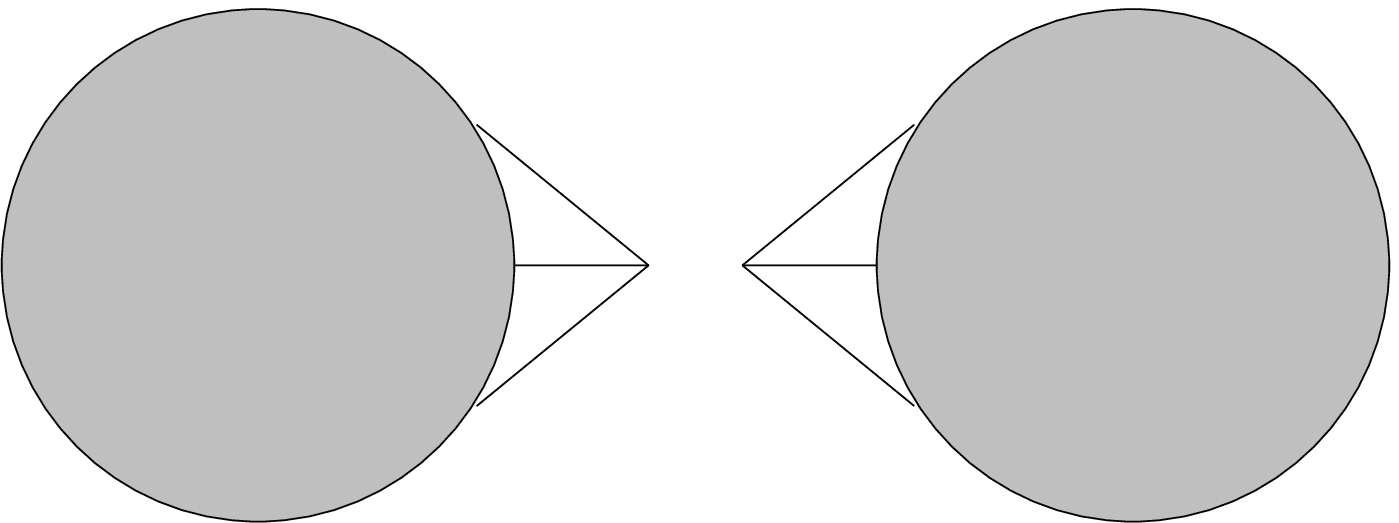}\end{array},
\ee
where again the filled circles represent generic spin networks.

Let us start with the zeroth order term in the master constraint. For the more generic state shown in FIG. \ref{fig:generic state} there are more possibilities to pick the contributions $i\frac{\sqrt{\Lambda}}{\tilde{n}}(2j_{\tilde{p}}+1)$ in order to get the first order of $|\phi>$ which give non-vanishing terms of the scalar product. In the following calculation we are going to pick one of these several possibilities. From the structure of the calculation it should be clear to the reader that the other alternatives give exactly the same result, namely
\baa
<\phi | \hat{C}^{(0)}  \triangleright |\Psi_+^{(1)}>&=&-\Lambda\int \left(\prod_{\tilde{h}} dg_{\tilde{h}}\right)\sum_{j_1,..., j_4,k}(2j_1+1)^2(2k+1)^2(2j_2+1)(2j_3+1)(2j_4+1)
\begin{array}{c}  \includegraphics[width=4cm,angle=360]{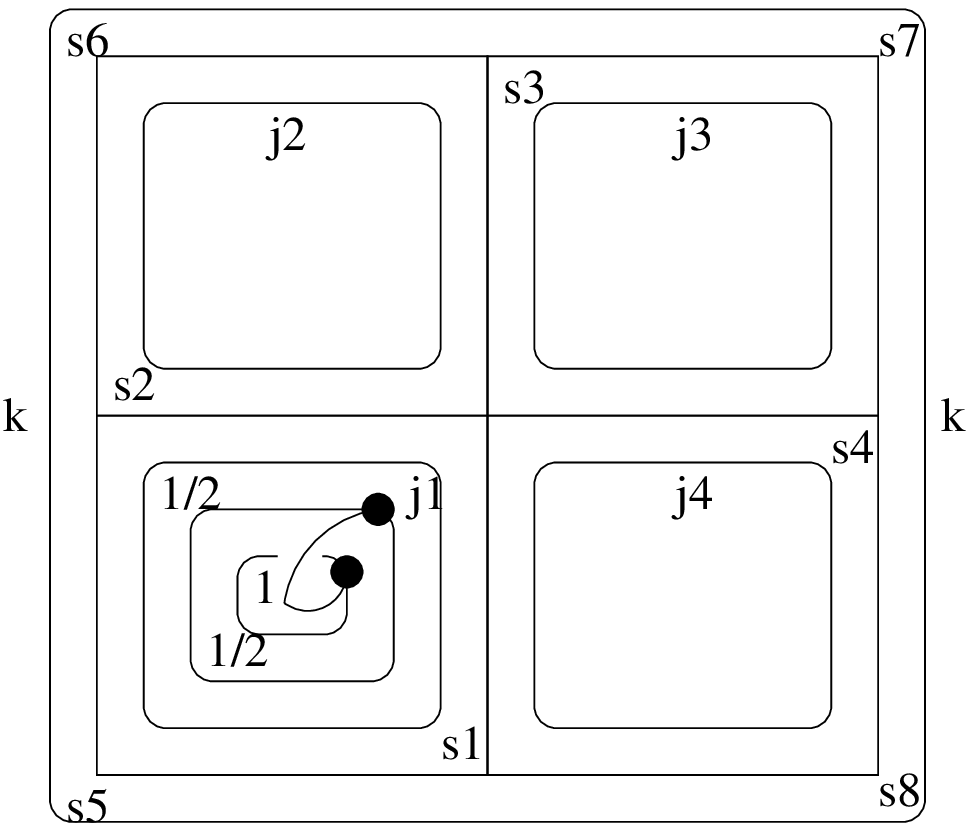}\end{array}\n\\
&=&-\Lambda\sum_{j_1,..., j_4,k}(2j_1+1)^2(2k+1)^2(2j_2+1)(2j_3+1)(2j_4+1)
\begin{array}{c}  \includegraphics[width=4cm,angle=360]{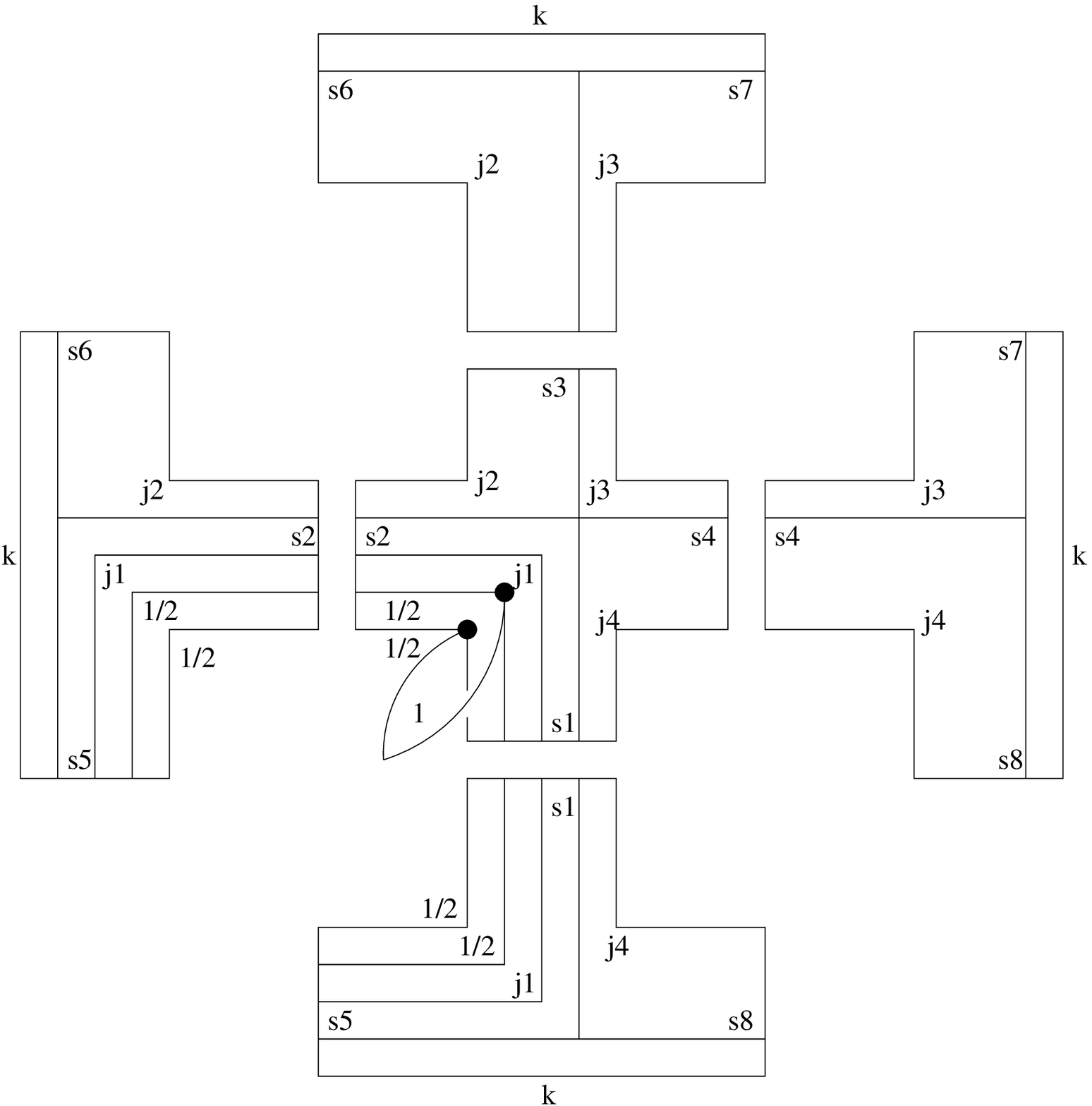}\end{array}\eaa
\ba
&=&-\Lambda \sum_{j_1,..., j_4,k}(2j_1+1)^2(2k+1)^2(2j_2+1)(2j_3+1)(2j_4+1)
\begin{array}{c}  \includegraphics[width=3.5cm,angle=360]{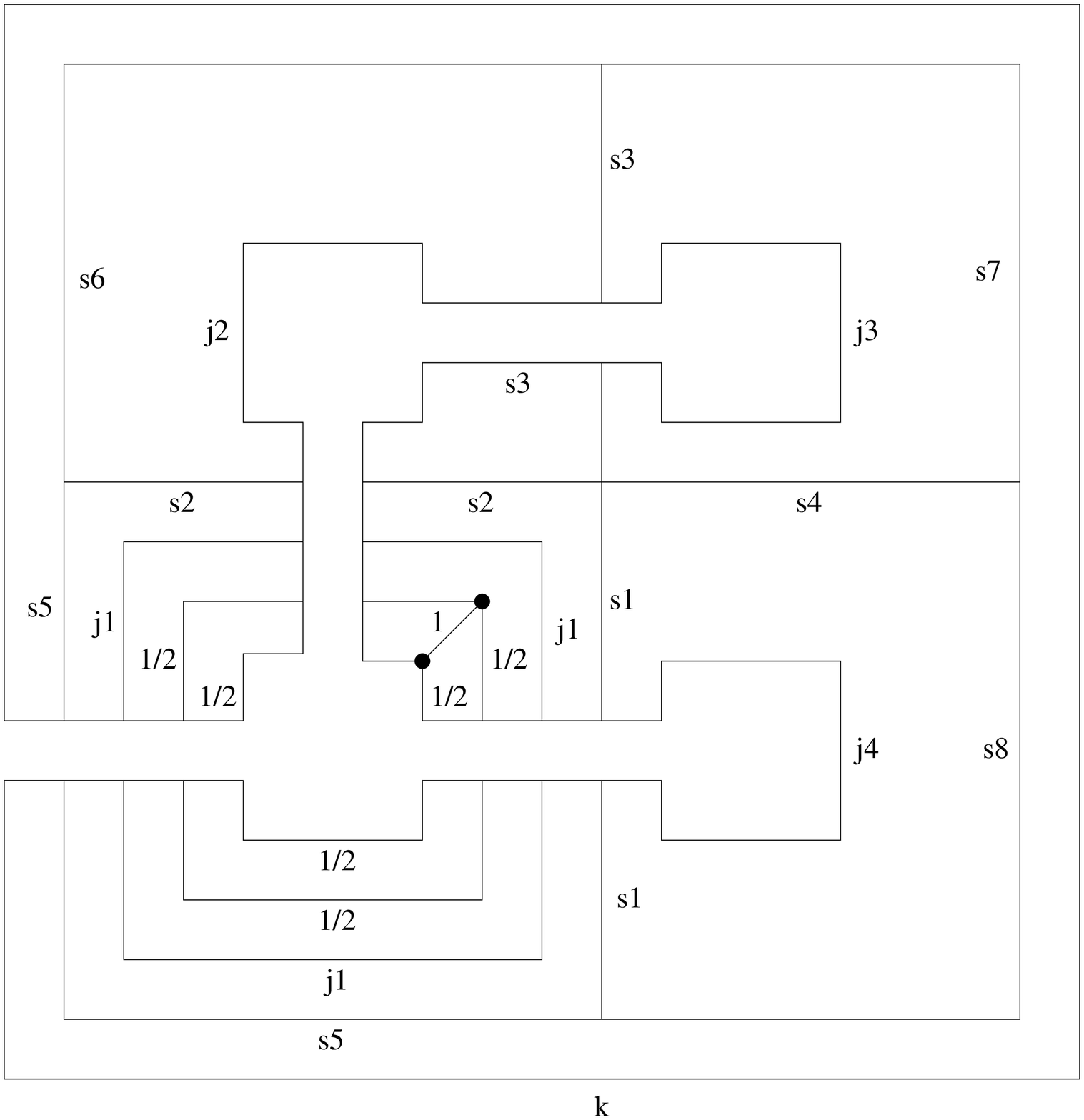}\end{array}\n\\
&=&-\Lambda \sum_{j_1, k}(2j_1+1)^2(2k+1)^2
\begin{array}{c}  \includegraphics[width=3.5cm,angle=360]{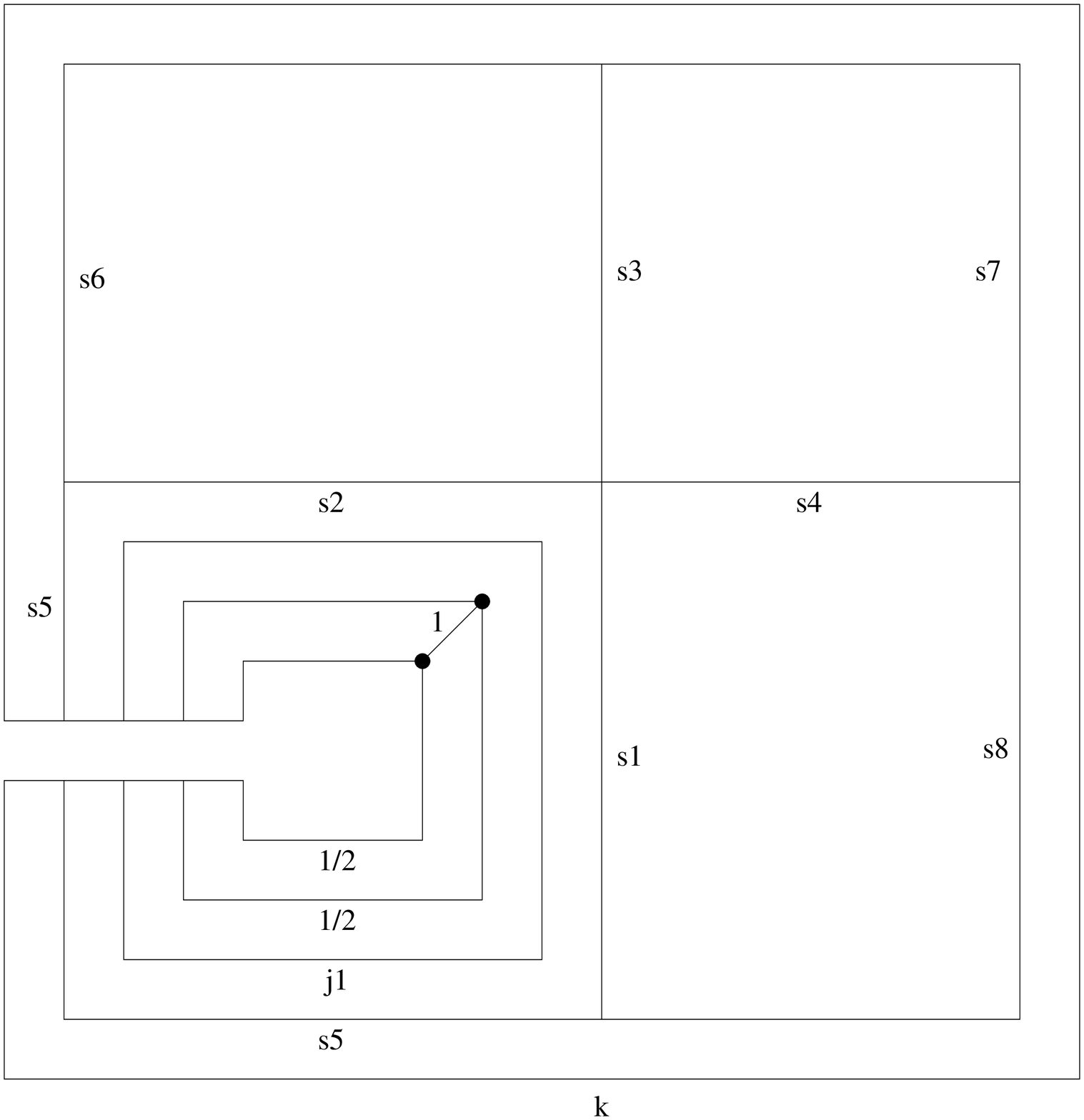}\end{array}
\n\\
&=&-\Lambda\left(\sum_i \frac{\sqrt{2i+1}}{2s_5+1}\left\{
\begin{array}{ccc}
s_2 & s_3 & i \\
s_7 & s_5 & s_6 \\
\end{array}
\right\}
\left\{
\begin{array}{ccc}
s_1 & s_4 & i \\
s_7 & s_5 & s_8 \\
\end{array}
\right\}\right)\sum_{j_1, k}(2j_1+1)^2(2k+1)^2
\begin{array}{c}  \includegraphics[width=3.5cm,angle=360]{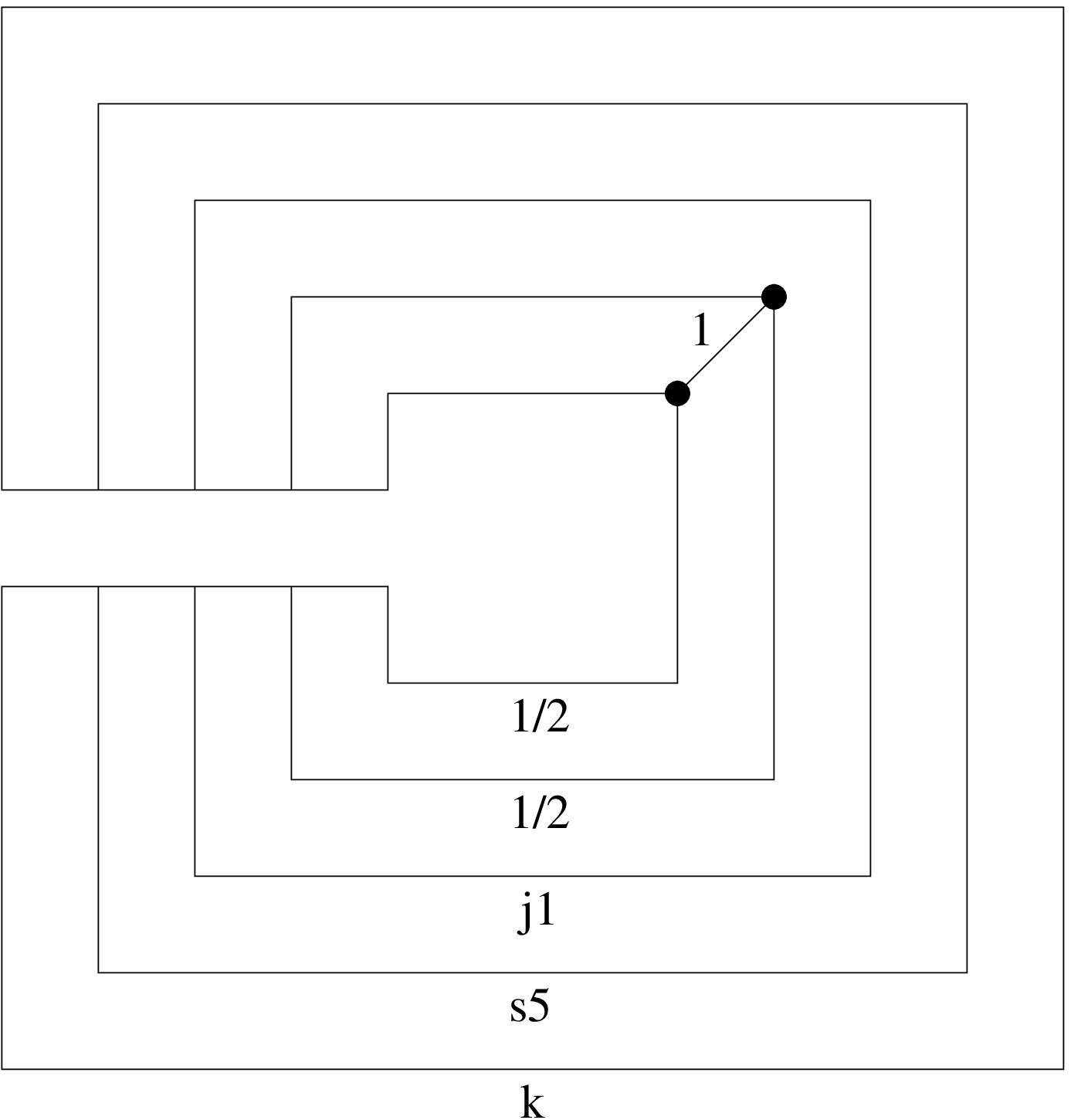}\end{array}\n\\
&=&2\Lambda\left(\sum_i \frac{\sqrt{2i+1}}{2s_5+1}\left\{
\begin{array}{ccc}
s_2 & s_3 & i \\
s_7 & s_5 & s_6 \\
\end{array}
\right\}
\left\{
\begin{array}{ccc}
s_1 & s_4 & i \\
s_7 & s_5 & s_8 \\
\end{array}
\right\}\right)
\sum_{j_1}(2j_1+1)^2(2s_5+1),\n\\\label{eq:ScalarGen1}
\ea
where in the last line we have used the previous result for a single loop state. A similar calculation shows that for the first order term in the master constraint we have
\ba
<\phi | \hat{C}^{(1)}  \triangleright |\Psi_+^{(0)}>&=&
4\Lambda\left(\sum_i \frac{\sqrt{2i+1}}{2s_5+1}\left\{
\begin{array}{ccc}
s_2 & s_3 & i \\
s_7 & s_5 & s_6 \\
\end{array}
\right\}
\left\{
\begin{array}{ccc}
s_1 & s_4 & i \\
s_7 & s_5 & s_8 \\
\end{array}
\right\}\right)\sum_{j_1,k}(2j_1+1)(2k+1)
 \begin{array}{c}  \includegraphics[width=2.8cm,angle=360]{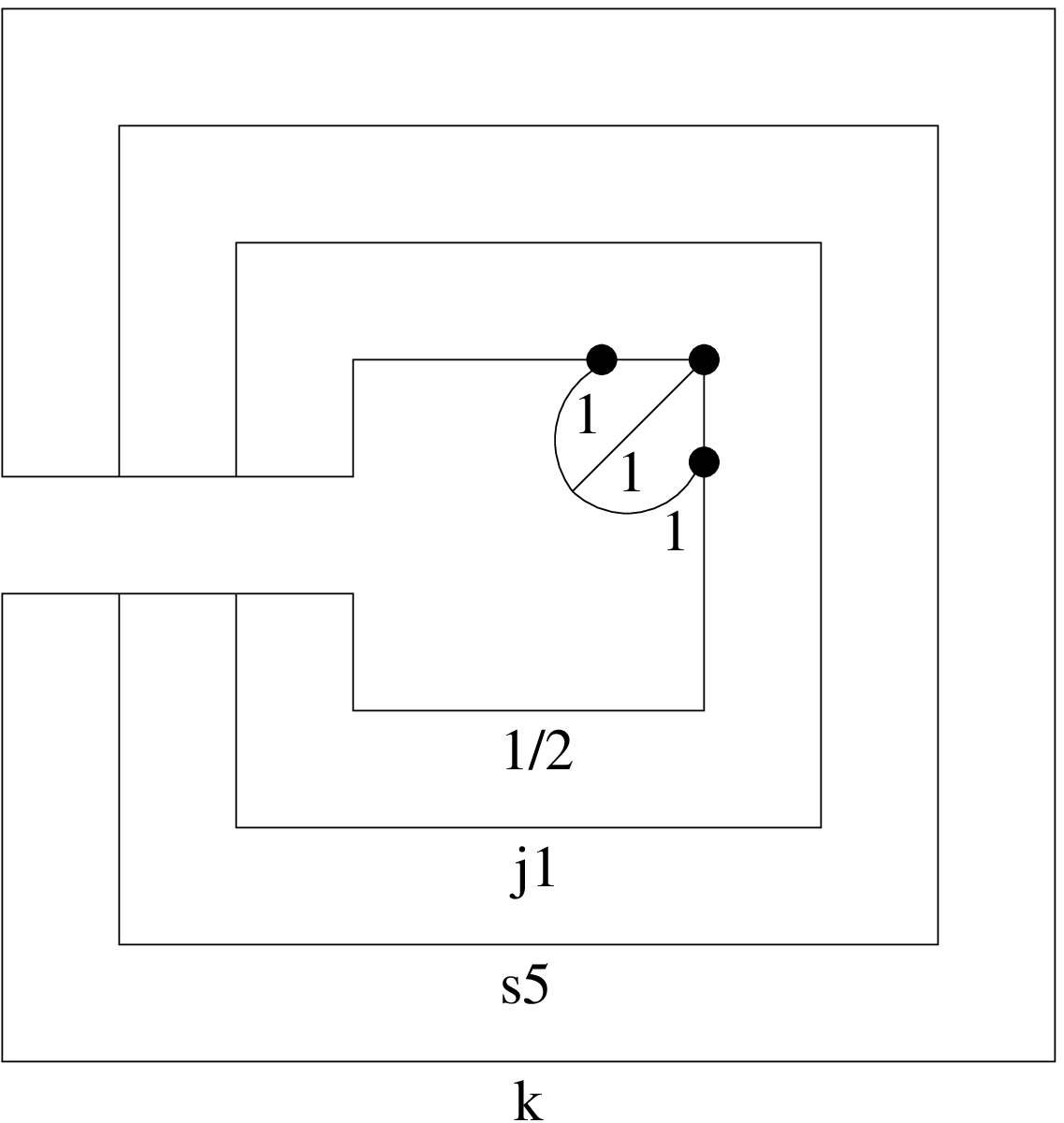}\end{array}\n\\
 &=&-2\Lambda\left(\sum_i \frac{\sqrt{2i+1}}{2s_5+1}\left\{
\begin{array}{ccc}
s_2 & s_3 & i \\
s_7 & s_5 & s_6 \\
\end{array}
\right\}
\left\{
\begin{array}{ccc}
s_1 & s_4 & i \\
s_7 & s_5 & s_8 \\
\end{array}
\right\}\right)\sum_{j_1}(2j_1+1)^2(2s_5+1),\label{eq:ScalarGen2}
\ea
where again we used the result of the calculation with a single loop state. Therefore, summing (\ref{eq:ScalarGen1}) and (\ref{eq:ScalarGen2}), it follows that the scalar product between the state shown in FIG. \ref{fig:generic state} and $(\hat{C}^2 \triangleright |\Psi>)^{(1)}$ vanishes. With a similar calculation it can be proven that
\be
<\phi|(\hat{C}^2 \triangleright |\Psi>)^{(1)}=0~~~~~~\forall \phi\in \Ha.
\ee
 This implies that our ansatz (\ref{eq:Ansatz}) is anihilated by the quantum version of the master constraint (\ref{eq:Master classic}) up to the first order in  $\Lambda$.

\section{Conclusions}\label{Conclusions}

The nature of the kinematical Hilbert space of LQG  is such that only variables of extended nature (holonomies and conjugate fluxes) can be quantized: in the kinematical LQG representation the fundamental operators representing phase space variables  and entering the definition of the quantum constraints of the theory need to be regularized. We have seen that, in the case of 2+1 gravity with a non-vanishing cosmological constant, this has the effect of introducing an anomaly in the regulated quantum constraints algebra, which is no longer associated to a structure Lie group.

On the other hand, other approaches to the same problem, already existing in the literature, strongly rely on the deformation of a Lie group: at the quantum level, in both the Turaev-Viro model and the Chern-Simons-Witten theory, the classical group gauge symmetry is replaced by a quantum group symmetry and observables expectation values are computed using the representation theory of the quantum group $U_q SL(2)$. Similarly, in the canonical approach to the combinatorial quantization formalism, the Hilbert space is constructed from the representation theory of the quantum group.

Therefore, even though the anomaly in the constraints algebra found in \cite{Anomaly} represents a serious obstacle to the implementation of the Dirac program, since the usual group averaging techniques used in the $\Lambda=0$ case cannot be applied anymore, it is an intriguing conjecture that it could at the end be related to a quantum group structure. 

In this paper, in order to provide evidence to this conjecture and carry on with the LQG program in the 2+1 context with $\Lambda>0$, I have relied on some alternative formulations developed for the quantization of systems with constraints algebras which are not associated to a structure Lie group. 
More precisely, in section \ref{Quantum Dim} I have introduced an ansatz for a physical state and shown that the scalar product, defined by the Ashtekar-Lewandowski mesure, between this state and a multiple loops state gives transition amplitudes in agreement with what one would expect from the Turaev-Viro model. 

In the last section of this paper, in order to show that the state previously introduced actually implements the right dynamicss of gravity, I have defined a master constraint for the system. Using the regularization scheme built in \cite{Anomaly}, I have then shown that the ansatz for the physical state solves the master constraint up to the first order in $\Lambda$.

Showing that the transition amplitudes of this state allow to recover the full Turaev-Viro model is a difficult task and it requires further investigation. Computing the higher orders of the amplitude $<\phi|\hat{C}^2 \triangleright |\Psi>$ is also an involved calculation, but it would be important to go at least to the second order to verify the validity of the proposal. On the other hand, the connection between the covariant approach to LQG and the Turaev-Viro model hasn't gone so far beyond the first order in $\Lambda$ (see, e.g., \cite{Freidel}). Work along these two directions is in progress and I hope that the picture presented here could become clearer in the nearby future.
Nevertheless, the results of this paper provide support to the expectation that the implementation of dynamics, in the case of a non-vanishing cosmological constant, should induce the appearance of a quantum group structure starting from the kinematical Hilbert space of LQG, where no quantum deformation of the Lie group is introduced by hand at any stage.

Recently, a different approach to the same problem has been studied in \cite{NPP}. In this work, the authors introduce a Chern-Simons connection which allows to rewrite the curvature constraint in presence of a cosmological constant in terms of Wilson loops of this non-commutative connection; as a first step towards the quantization of this constraint, \cite{NPP} studies the canonical quantization of the holonomy of such a connection on the kinematical Hilbert space of loop quantum gravity, providing an explicit construction of the quantum holonomy operator. The authors find a close
relationship between the action of the quantum holonomy at a crossing and Kauffman's $q$-deformed crossing identity, which, together with the quantum dimension, is the fundamental ingredient for the definition of the Turaev-Viro model \cite{KL}. Therefore, while following different approaches, both \cite{NPP} and the present work are inspired by the same philosophy (no introduction of quantum group structures by hand) and use the same LQG technology, obtaining results which can somehow be seen as complementary and narrowing the gap between the covariant and canonical approaches to the problem of $2+1$ quantum gravity in the presence of a non-vanishinng cosmological constant.

\section{Acknowledgements}

I am very grateful to Alejandro Perez for providing insightful criticism and his support during the elaboration of this manuscript. 
This work was supported by {\em Marie Curie} EU-NCG network.

\end{document}